# Liquid Phases in SU3 Chiral Perturbation Theory $(SU3\chi^{PT})$: Drops of Strange Chiral Nucleon Liquid $(S\chi NL)$ & Ordinary Chiral Heavy Nuclear Liquid $(\chi NL)$

Bryan W. Lynn

University College London, London WC1E 6BT, UK
&
Theoretical Physics Division, CERN, Geneva CH-1211
bryan.lynn@cern.ch

## Abstract

$SU3_L \, xSU3_R$ chiral perturbation theory ($SU3\chi PT$) identifies hadrons (rather than quarks and gluons) as the building blocks of strongly interacting matter at low densities and temperatures. It is shown that $SU3\chi PT$ of nucleons and kaons <u>simultaneously</u> admits two chiral nucleon liquid ($\chi NL$) phases at zero external pressure with well-defined surfaces: 1) ordinary chiral heavy nuclear liquid drops and 2) a new electrically neutral Strange Chiral Nucleon Liquid ($S\chi NL$) phase with both microscopic and macroscopic drop sizes.

Analysis of chiral nucleon liquids is greatly simplified by an $SU3\chi PT$ spherical representation and identification of new class of solutions with $\partial_\mu \hat{\pi}_a = 0$. $\chi NL$ vector and axial-vector currents obey all relevant CVC and PCAC equations. $\chi NL$ are therefore solutions to the $SU3\chi PT$ semi-classical liquid equations of motion. Axial-vector chiral currents are conserved inside macroscopic drops of $S\chi NL$, a new form of baryonic matter with zero electric charge density, which is by nature dark.

The numerical values of all $SU3\chi PT$ chiral coefficients (to order $\Lambda^0_{\chi SB}$ up to linear in strange quark mass $m_S$) are used to fit scattering experiments and ordinary chiral heavy nuclear liquid drops (identified with the ground state of ordinary even-even spin-zero spherical closed-shell nuclei). Nucleon point coupling exchange terms restore analytic quantum loop power counting and allow neutral $S\chi NL$ to co-exist with chiral heavy nuclear drops in $SU3\chi PT$ (i.e. using these same numerical values, without new adjustable parameters). Due primarily to experimentally measured order $\Lambda^0_{\chi SB}$ chiral coefficients in the kaon-nucleon explicit symmetry breaking sector, $S\chi NL$ tends to high baryon number density and chemical potential, but these can both be significantly lowered by higher order $\Lambda^{-1}_{\chi SB}$ explicit symmetry breaking terms.

For certain chiral coefficients, finite microscopic and macroscopic drops of $S\chi NL$ may be the ground state of a collection of nucleons. Thus, ordinary heavy nuclei may be meta-stable, while oceans of $S\chi NL$ may force qualitative and experimentally observable changes to the neutron star equation of state.

## Section 1: Introduction and List of Main New Results

What is the ground state of QCD at zero external pressure with non-zero baryon number? $SU3\chi PT$, based entirely on the global symmetries of QCD, identifies hadrons (rather than quarks and gluons) as the building blocks of strongly interacting matter at very low densities and temperatures [1,2,3]. In doing so, it simply acknowledges (as a starting point) a still-mysterious experimental fact: Nature first makes hadrons and then makes everything else, like ordinary nuclei [4,5,6], out of them.

$SU3\chi PT$ (effective Lagrangian) power counting [3] includes all analytic higher order quantum loop corrections into tree-level amplitudes. The resultant perturbation expansion in $1/\Lambda_{\chi SB}$ with (chiral symmetry breaking scale) $\Lambda_{\chi SB} \approx 1 GeV$ renders $SU3\chi PT$'s strong interaction predictions calculable in practice! Its low-energy dynamics of baryon and pseudo-Goldstone octet is our best understanding of the experimentally observed low-energy dynamics of QCD strong interactions: Pseudo-Goldstone masses; Soft pion and kaon scattering; Applicability of SU2 and SU3 current algebra; Vector CVC and axial-vector PCAC; Semi-leptonic K decay; Baryon masses (Gell-Mann Okubo); Hyperon non-leptonic decay; Hyperon magnetic moments; Baryon axial couplings; Non-leptonic K decay; Zweig rules; Goldberger-Treiman relation, etc. It has also been demonstrated [5,6] that $SU2 \times U1\chi^{PT}$ of nucleons and pions explains the detailed structure of very light nuclei (e.g. deuteron). Significant progress has also been made in explaining the detailed structure of the ground state of certain heavy nuclei [8,9,10,11,12].

B.W.Lynn, A.E.Nelson and N.Tetradis (LNT) [13] proposed and demonstrated the (mathematical) existence of microscopic and macroscopic drops of electrically neutral "Strange Chiral Nucleon Liquids" ($S\chi NL$) in $SU3\chi PT$ of neutrons and neutral kaons, but did not find ordinary nuclear matter there. B.Ritzi & G.Gelmini [14] further developed $S\chi NL$ (i.e. our new hypothetical state) but did not simultaneously find ordinary nuclear liquids (i.e. the ordinary observed states), generating a real credibility problem for $S\chi NL$ in $SU3\chi PT$. We show here that $S\chi NL$ can co-exist, (with the same chiral coefficients) with ordinary nuclear liquids within $SU3\chi PT$, significantly increasing the credibility of both $S\chi NL$ and $SU3\chi PT$. This co-existence is a requirement for legitimate comparison between them and allows us to question whether $S\chi NL$ might be the ground state of a collection of nucleons, where ordinary heavy nuclei are only meta-stable. It also legitimizes the question of whether oceans of $S\chi NL$ force qualitative and experimentally observable changes to the equation of state and surface of neutron stars.

The constituents of strange $SU(3)_L \times SU(3)_R$ baryonic matter [15,13] are taken to be hadrons: i.e. baryons and pseudo-Goldstone bosons transforming as octets (for simplicity we ignore higher mass hadronic chiral multiplets). It was called "Strange Baryon Matter" ($SBM$) to distinguish it from "Strange Quark Matter", whose constituents are presumed to be quarks. We find $SBM$ naming conventions useful in distinguishing its various forms: "B"="Baryon Octet", "N"="Nucleons", "S"="Strange", "G"="Gases/Plasmas", "L"="Liquids", "C"="Crystals/Solids",

"$\chi$"="$SU3\chi PT$". Then: "Strange Baryon Gases/Plasmas" ($SBG$) disperse to baryons in zero external pressure, but may form in the interior of neutron stars (i.e. held together with gravity) or in the early universe [15]; "Strange Baryon Liquids" ($SBL$) [13] are liquid drops (with well-defined surfaces) of $SBM$ at zero external pressure (i.e. not crystalline or gaseous) where it costs energy to either increase or decrease density [13,14]; "Strange Baryon Crystals or Solids" ($SBC$) may also exist; In "Strange Nucleon Matter" ($SNM$) baryons are primarily protons and neutrons and comes as $SNC, SNL$ and $SNG$ [15,16]. LNT constructed (non-topological soliton) "Strange Nucleon Liquid" ($SNL$) Q-Balls; "Strange Chiral Baryon Matter" ($S\chi BM$) is $SBM$ derivable from $SU3\chi PT$; Non-strange versions of these simply drops the prefix "S": e.g. we argue that ordinary nuclear liquid drops are derivable from $SU3\chi PT$ and are therefore a "Chiral Nucleon Liquid" ($\chi NL$); In this paper, we study electrically neutral $S\chi NL$ and charged $\chi NL$. Any (strange or non-strange) liquid phase of $SU3\chi PT$ will simply be called a "Chiral Liquid".

Central to $SNL$ analysis (and simplification) is the "LNT $nK^0\overline{K^0}$ Ansatz" [13]:
1. An $SU(3)_{L+R}$ singlet field $\theta$ (the magnitude of the pseudo-Goldstone field, rather than its direction $\hat{\pi}_a$ in SU3) is identified;
2. Species: Baryon $B \to n$ (neutrons), Goldstones $\to K^0 = \overline{K^0} = \frac{f_\pi}{\sqrt{2}}\theta$ [16];
3. This and $\pi^0, \pi^+, \pi^-, K^+, K^-, \eta = 0$ are <u>shoved directly</u> into the $SU3\chi PT$ Lagrangian. This is famously (mathematically) dangerous in quantum field theory. For example, isospin currents $T_a^\mu, T^{5\mu}{}_a$ $a=1,2,3$ arise from the variation of $\pi^0, \pi^+, \pi^-$ fields (set to zero in the butchered Lagrangian!), while hypercharge currents $T_8^\mu, T^{5\mu}{}_8$ arise from variation of the $\eta$ field (also set to zero);
4. Although $\partial_0\theta = 0$ ensured Conservation of Vector Currents (CVC) for the $T^\mu_{Strangeness}$ current, CVC equations for $T_a^\mu$ $a=1,2,3,4,5,6,7$ were ignored;
5. The Partial Conservation of Axial-vector Currents (PCAC) was ignored for all currents $T^{5\mu}{}_a$ $a=1,2,3,4,5,6,7,8$;
6. For fixed large baryon number, the nucleons are replaced by a perfect fluid in the Thomas-Fermi mean field approximation: $\theta$ is assumed to vary sufficiently slowly in space so as to treat the nucleons as plane waves moving in its self-consistent background condensate field. The fermions are stacked, with Hartree interactions (Hartree-Fock here!), into a Fermi sea with space-dependent spherically symmetric Fermi-momentum $k_F(r)$ and constant baryon number chemical potential $\mu_B$;
7. Variation of $\theta$ in the Lagrangian gives a semi-classical 2$^{nd}$ order partial differential equation of motion. A spherically symmetric S-wave $\theta(r)$ then obeys a 2$^{nd}$ order ordinary differential equation (ODE);
8. In recognition of these dangers, LNT called the above an "ansatz", to distinguish it from a "solution".

We will show here that $S\chi NL$ [13,14], $SNL$ [13] and ordinary chiral heavy nuclear liquids [8,9,12] satisfy all relevant $SU3\chi^{PT}$ CVC and PCAC equations in the liquid phase and (having avoided the dangers listed above) are solutions of the tree level semi-classical liquid equations of motion (i.e. rather than just an ansatz).

The LNT $nK^0\overline{K^0}$ solution vastly simplifies $SNL$ analysis: Both axial vector and vector coupling terms between baryons and pseudo-Goldstones vanish identically

$$FTr\overline{B}\gamma^\mu\gamma^5[A_\mu,B] + DTr\overline{B}\gamma^\mu\gamma^5\{A_\mu,B\} = 0; \quad iTr\overline{B}\gamma^\mu[V_\mu,B] = 0; \quad (1)$$

Nucleon-pseudo-Goldstone (spontaneous breaking) forces in $L_{\chi_{PT}}^{Symmetric}$ play no role; Neutron $J_{Baryon}^\mu$ is automatically conserved. The only $SU(3)_L \times SU(3)_R \chi^{PT}$ sector driving $SNL$ binding energies are <u>explicit</u> chiral symmetry breaking terms [17]

$$L_{\chi_{PT}}^{\pi;SymmetryBreaking} \to -U^K = -2m_K^2 f_\pi^2 s_{\theta/2}^2$$
$$L_{\chi_{PT}}^{B;SymmetryBreaking} \to 2m_S(2a_3 + a_2 + a_1)\overline{n}n s_{\theta/2}^2 \quad (2)$$

from the kaon mass and the attractive $\sigma_{\pi N}$ force between nucleons and S-wave $\theta$ condensate respectively. We use the notation $s_\theta = \sin(\theta), c_\theta = \cos(\theta)$ throughout this paper. The search for, proof of existence/uniqueness, construction and display of semi-classical solutions $\theta(r)$ is simplified because its ODE can be mapped onto Newtonian potential motion and "Roll-around-ology" [17,18,14,19,15]: i.e. a non-relativistic point particle moves a "distance" $\theta(\tilde{r})$, in "time" $\tilde{r} = 2f_\pi r$, with "friction" $\frac{2}{\tilde{r}}\frac{d\theta}{d\tilde{r}}$, in a "Newtonian potential" $V_{Newton} = \frac{(P^\Psi - U^K)}{4f_\pi^4}$, with $P^\Psi$ the nucleon Thomas-Fermi liquid pressure.

LNT explicitly showed the existence and properties of $SNL$ in 2 very different theories: 1) $SU3\chi^{PT}$ and 2) non-linear $SU(3)_L \times SU(3)_R$ coupled to a model of ordinary nuclear matter. Spherical Q-Balls (non-topological solitons) displayed had: Any desired baryon number (above some minimum); A saturating "liquid" interior for large baryon number; Baryon number density and condensate ($n^\dagger n, \theta$) almost constant out to a large radius R for large baryon number; ($n^\dagger n, \theta$), becoming independent of total baryon number in the large baryon limit ($\infty SNL$); A thin surface $\Delta R \approx f_\pi^{-1}$ with ($n^\dagger n, \theta$) dropping to zero in the external vacuum; Strangeness to baryon number $\approx (1-c_\theta)$; Microscopic or macroscopic radius; In some cases, strange Q-Balls had binding energy/nucleon significantly deeper than ordinary nuclear matter and Strange Q-Stars (non-topological soliton stars) called $S\chi NLStar$ were also shown to exist.

LNT's non-linear $SU(3)_L \times SU(3)_R$ symmetric theory coupled (for the 1$^{st}$ time) $SU3\chi^{PT}$ to a Walecka-type model of ordinary heavy nuclei and simultaneously contained <u>both</u> strange and non-strange nucleon liquids: i.e. $SNL$ <u>co-existed</u> with symmetric bulk nuclear matter. In reliance on the empirically successful and vast literature surrounding Walecka's nuclear models [20,21], this theory contained microscopic non-strange nuclear liquid drops (with mathematical solutions emerging as a species of fermion Q-Ball [18] or non-topological soliton [22]) which are identified with the ground state of ordinary even-even spin-zero spherically symmetric heavy nuclei (e.g. $_{20}^{20}Ca_{40}, _{50}^{40}Zr_{p0}, _{126}^{82}Pb_{208}$). $SNL$ densities were lowered to levels acceptable to $SU3\chi^{PT}$ by explicit 4-fermion point-coupling vector repulsion and implicit scalar attraction 4-fermion point-coupling and long range surface terms which are usual in Walecka models.

LNT's $SU3\chi^{PT}$ model [13] is the 1st instance of $S\chi NL$ chiral non-topological solitons. But because 4-fermion point-coupling interactions $L_{\chi PT}^{4-B,Symmetric}$ were ignored, ordinary bulk nuclear matter and finite nuclear liquid drops did not emerge. Worse, the $S\chi NL$ baryon number densities found were too high $\Psi^{\dagger}\Psi \approx 10.8[\Psi^{\dagger}\Psi]_{NuclearMatter}$. The authors complained that, at such high densities, both $SU3\chi^{PT}$ and the baryon and meson description of nuclear matter should break down: e.g. a valid description of matter at such high densities might more probably involve quarks and gluons.

B.W.Lynn showed [19] that the baryon number density of <u>non-strange</u> chiral nucleon liquids ($\chi NL$), in Isospin-Hypercharge $SU2 \times U1\chi^{PT}$, was lowered to acceptable levels by the introduction of liquid (order $\Lambda_{\chi SB}^0$) 4-nucleon (protons and neutrons) point-coupling (including nucleon-exchange) terms. The coefficient of the attractive $\sigma_{\pi N}$ term in $L_{\chi PT}^{B,SymmetryBreaking} \to 2\beta^{\Psi\pi}\sigma_{\pi N}\overline{\Psi}\Psi s_{\theta/2}^2$ (nucleons $\Psi = \begin{bmatrix} p \\ n \end{bmatrix}$ and S wave $\pi^0\pi^0$ condensate $\theta$) was artificially increased from its experimental value $\beta^{\Psi\pi} \sim 1$ to $\beta^{\Psi\pi} \sim 6.6$. G.Gelmini, B.W.Lynn & B.Ritzi [23] therefore hypothesized that supplementing the LNT $nK^0\overline{K^0}$ ansatz with 4-neutron point-coupling terms (and experimentally large $\beta^{nK} \approx 9.28$) would lower baryon number densities of electrically neutral "Infinite Strange Chiral Nucleon Liquids" $\infty S\chi NL$ to acceptable $SU3\chi^{PT}$ levels.

B.Ritzi & G.Gelmini showed [14] that this hypothesis was true and found $\infty S\chi NL$ with: Acceptable (i.e. where $SU3\chi^{PT}$ may be applicable) baryon number densities $\Psi^{\dagger}\Psi \approx (5-6)[\Psi^{\dagger}\Psi]_{NuclearMatter}$; Binding energy/neutron (in some cases) significantly deeper than ordinary nuclear matter; An important (graphical) constraint which partially ensures that soliton solutions do not encounter a multi-valued Newtonian potential; Important effects of certain higher order $\Lambda_{\chi SB}^{-1}$ symmetry breaking terms which lowered baryon number densities and chemical potentials even further. But, since point-coupling exchange terms were absent, ordinary chiral nuclear liquids did not co-exist with their $\infty S\chi NL$.

### 1.2 Main New Results of This Paper

We show here that (to order $\Lambda_{\chi SB}^0$ up to linear in $m_S$) in $SU3\chi PT$:
1. Solutions to the semi-classical equations of motion of nucleons and kaons <u>simultaneously</u> admit two liquid phases (together with finite drops) which exist in vacuum and do not require external pressure (e.g. gravity):

- Ordinary finite chiral nuclear liquid drops ($\chi NL$), identified as the ground state of ordinary even-even spin-zero spherical closed-shell heavy nuclei;
- An (as yet unobserved) electrically neutral liquid (nucleons + kaon condensate) phase - Strange Chiral Nucleon Liquids ($S\chi NL$) - with both macroscopic and microscopic drop size. Drops of $S\chi NL$ are true liquids (i.e. not crystalline or gaseous) with well-defined surfaces;

2. Analysis of chiral liquids is vastly simplified by
   - Introduction of a spherical representation $\pi_a = f_\pi \hat{\pi}_a \theta; \quad \hat{\pi}_a \hat{\pi}_a = 1$;
   - Identification of a new class of solutions $\partial_\mu \hat{\pi}_a = 0$;

3. Vector and axial vector currents, CVC and PCAC:
   - Both ordinary finite chiral nuclear liquid drops and $S\chi NL$ obey all relevant vector current CVC and axial vector current PCAC relations in the liquid with $\partial^0 \theta = 0$, and are therefore to be regarded as solutions of the semi-classical liquid equations of motion;
   - Non-trivial PCAC constraints (i.e. $\Pi_3^{\bar{\Psi}\Psi}, \Pi_8^{\bar{\Psi}\Psi} = 0$) on ordinary chiral nuclear liquid and $S\chi NL$ axial vector currents $T_3^{5\mu}, T_8^{5\mu}$ are satisfied: certain nucleon bi-linears (which re-introduce the quantum numbers of $\pi^0, \eta$ and partially arise from 4-nucleon contact terms) vanish in chiral liquids;
   - Higher order $\Lambda_{\chi SB}^{-1}, \Lambda_{\chi SB}^{-2}$ corrections crucial to <u>current</u> high accuracy nuclear Skyrme models are invariant under local $SU3\chi^{PT}$ transformations, do not contribute to $SU3_{L+R}$ or $SU3_{L+R}$ currents and satisfy CVC and PCAC;

4. Chiral axial-vector currents are conserved inside macroscopic $S\chi NL$ drops:

$$\left[\partial_\mu (T_6^{5\mu} \pm iT_7^{5\mu})\right]_{\infty S\chi NL} = -f_\pi^2 \left[\partial^2 \theta\right]_{\infty S\chi NL} = -\left[\frac{\partial}{\partial \theta} L_{\chi PT}^{SymmetryBreaking}\right]_{\infty S\chi NL} = 0; \quad (3)$$

5. An improved search method for solutions to the $SU3\chi^{PT}$ semi-classical equations of motion makes use of 1, 2, 3, 4 above. Since the mathematics of liquid drops in $SU3\chi^{PT}$ maps self-consistently (including 4-nucleon point-coupling terms) onto Newtonian potential motion, its existence/uniqueness properties (known since antiquity) can be used to classify (strange and non-strange) $\chi NL$ according to the existence (i.e. versus non-existence) of Q-Ball non-topological solitons (in the phase space of baryon chemical potential and number density): i.e. Roll-around-ology is also a powerful systematic search method to find those circumstances in which solutions <u>do not</u> exist.

6. In the interior of macroscopic $\infty S\chi NL$ drops:
   - The scalar density is constant and the algebra is further simplified

$$\left[\bar{n}n\right]_{\infty S\chi NL} = \frac{f_\pi^2 m_K^2}{m_S (2a_3 + a_2 + a_1)} \quad (4)$$

   - Therefore, due to experimentally measured parameters in the kaon explicit symmetry breaking sector [17], $\infty S\chi NL$ has high baryon number densities

$$\frac{\left[n^\dagger n\right]_{\infty S\chi NL}}{\left[\Psi^\dagger \Psi\right]_{NuclearMatter}} \geq \frac{\left[\bar{n}n\right]_{\infty S\chi NL}}{\left[\Psi^\dagger \Psi\right]_{NuclearMatter}} \approx 2.593; , \quad (5)$$

explaining why raising $a_3$ (via its $\pm 0.2$ experimental errors) lowers baryon number density [14];

7. Finite microscopic and macroscopic drops of $S\chi NL$ and 4-nucleon point-coupling exchange terms:
   - The numerical values of order $\Lambda_{\chi SB}, \Lambda^0_{\chi SB}$ chiral coefficients (up to linear in $m_S$) are all used here to fit scattering experiments and ordinary nuclear liquid drops, interpreted as the ground state of ordinary even-even spin-zero spherical closed-shell heavy nuclei,: (e.g. $^{20}_{20}Ca_{40}, ^{40}_{50}Zr_{p0}, ^{82}_{126}Pb_{208}$);
   - Nucleon point-coupling exchange terms are introduced (for correct $\Lambda^0_{\chi SB}$ quantum loop power counting) and are identified as the crucial ingredient allowing neutral $S\chi NL$ to co-exist (with ordinary heavy nuclei) in $SU3\chi PT$ with the same set of chiral coefficients (i.e. with no new adjustable parameters);
   - The simultaneous fit to ordinary chiral nuclear liquid drops implies a significant gain in credibility for the graphical and mathematical tools used in the construction of $S\chi NL$ solutions;
   - Only since $S\chi NL$ and ordinary chiral nuclear liquids are shown to co-exist in $SU3\chi PT$ is comparison between them legitimate and possible: e.g. comparing the binding energy per nucleon of finite macroscopic drops of $S\chi NL$ with the ancient Weizsacker empirical nuclear mass formula [24];
8. We construct finite microscopic Q-Ball (non-topological soliton) solutions;
9. Numerical Results:
   - $\infty S\chi NL$ has high baryon number density and chemical potential:
   $$\frac{[n^\dagger n]_{\infty S\chi NL}}{[\Psi^\dagger \Psi]_{NuclearMatter}} \sim 4.84 - 6; \quad [\mu^n_B]_{\infty S\chi NL} \geq 0.896 GeV; \qquad (6)$$
   - Although we agree with certain qualitative conclusions of B.Ritzi & G.Gelmini [14], we disagree with their numerical results and identify an error. If this error is algebraic (rather than typographical), their results (in principle) mis-balance point-coupling repulsion, destroy relativistic covariance of their Thomas-Fermi liquid, violate PCAC and fail to conserve chiral axial-vector currents in the $\infty S\chi NL$ interior;

Including certain (cherry-picked) $SU3\chi PT$ higher order $\Lambda^{-1}_{\chi SB}$ terms linear in $m_S$ (from the explicit kaon-nucleon symmetry breaking sector) into $S\chi NL$:

10. Vector and axial vector currents are unaffected;
11. CVC and PCAC remain satisfied;
12. In the $\infty S\chi NL$ interior, the baryon scalar density is again constant (see Equation 76), resulting in further simplification of the algebra.
13. Baryon scalar densities can be significantly lowered (again, due to the kaon explicit symmetry breaking sector)
$$\frac{[\bar{n}n]_{\infty S\chi NL}}{[\Psi^\dagger \Psi]_{NuclearMatter}} = 2.113; \quad (C^S_{201} - \overline{C^S_{210}}) = -1.5; \quad a_3 = 1.5; \qquad (7)$$
14. Baryon number densities and chemical potentials can be significantly lowered
$$\frac{[n^\dagger n]_{\infty S\chi NL}}{[\Psi^\dagger \Psi]_{NuclearMatter}} \sim 3.36 - 6; \quad [\mu^n_B]_{\infty S\chi NL} > 0.807 GeV; \qquad (8)$$

in qualitative agreement with Reference [14], although we disagree with their numerical results;

15. Two new constraints (i.e. for fixed $\left[n^\dagger n\right]_{\infty S\chi NL}$), are derived
$$m^N > \left[\mu_B^n\right]_{\infty S\chi NL}; \quad m^N > \left(\left[\mu_B^n\right]_{\infty S\chi NL}\right)_{Maximum} \geq \left[\mu_B^n\right]_{\infty S\chi NL}); \qquad (9)$$
These place lower limits on $(-C_{200}^S + \overline{C_{200}^S})$ and upper limits on $(C_{200}^V - \overline{C_{200}^V})$ and deeply constrain $\infty S\chi NL$;

16. We construct finite microscopic Q-Ball (non-topological soliton) solutions;
17. For certain chiral coefficients, $S\chi NL$ may be the ground state of a collection of nucleons: i.e. ordinary chiral heavy nuclei may only be meta-stable;

Neutron Stars

18. Using approximate equations of state (EOS) in the construction of $S\chi NLStars$ [13,25], a chiral form of Q-Star [18,19], we argue that oceans of $S\chi NL$ may force qualitative and experimentally observable changes to neutron star EOS.

This paper is organized as follows: Section 2 introduces a spherical representation of $SU3\chi PT$ with baryon & pseudo-Goldstone boson octets and focuses attention on solutions with $\partial_\mu \hat{\pi}_a = 0$; Section 3 develops $SU3\chi PT$ 4-nucleon point-coupling (including fermion exchange) terms; Section 4 addresses Thomas–Fermi mean field liquids and proves that all relevant CVC and PCAC relations are satisfied in chiral liquids; Section 5 develops ordinary non–strange $SU(2) \times U(1) \chi^{PT}$ nuclear liquid drops ($\chi NL$) with $\theta = 0$, relating them to nuclear Skyrme models and the ground state of symmetric ($Z = N$) and asymmetric ($N \neq Z$) even-even spin-zero spherical closed-shell heavy nuclei; Section 6 develops electrically neutral $S\chi NL$ (with $\theta \neq 0$) in $SU3\chi^{PT}$ and focuses on macroscopic saturating solutions $\infty S\chi NL$. We give an algebraic version of a previous graphical [14] upper limit on $(-C_{200}^S + \overline{C_{200}^S})$ and lower limit on $(C_{200}^V - \overline{C_{200}^V})$. We derive new lower limits on $(-C_{200}^S + \overline{C_{200}^S})$ and upper limits on $(C_{200}^V - \overline{C_{200}^V})$, include certain higher order terms, and make comparison with previous results [13,14]; Section 7 is on finite and microscopic neutral $S\chi NL$ liquid drops; Section 8 discusses $S\chi NL$ neutron stars ($S\chi NLStar$); Section 9 discusses ideas going forward. Because this paper is aimed at a very broad audience (theoretical and experimental nuclear, high-energy and astrophysicists), we collect in Appendices 1 and 2 useful results and formulae from $SU3\chi^{PT}$ and Thomas-Fermi mean field liquids respectively; Appendix 3 writes $\Pi_a^{\overline{\Psi}\Psi}$ and $\Pi_8^{\overline{\Psi}\Psi}$ in terms of the usual nucleon bi-linears; Appendix 4 gives details on the spherical representation of $SU3\chi^{PT}$, solutions with $\partial_\mu \hat{\pi}_a = 0$, vector and axial vector currents and satisfaction of CVC and PCAC in chiral liquids (i.e. both $S\chi NL$ and ordinary nuclear liquid drops); Appendix 5 gives further order $\Lambda_{\chi SB}^0$ results. Appendix 6 gives details on our algebraic version of a partial graphical constraint [14] from Roll-around-ology. Appendix 7 shows how a program of systematic improvement of higher order (to $\Lambda_{\chi SB}^{-2}$) Skyrme mean field $\theta = 0$ nuclear models, together with nuclear experiments, could solidify the status of ordinary heavy nuclei and nuclear liquids as $SU(2) \times U(1) \chi^{PT}$ semi-classical solutions and strongly improve constraints on $S\chi NL$; Appendix 8 gives a new constraint from Roll-around-ology and shows that conservation of axial-vector currents inside $\infty S\chi NL$, in the presence of certain $\theta \neq 0$ higher order ($\Lambda_{\chi SB}^{-3}$) order symmetry-breaking point-contact (with nucleon-exchange) terms, can significantly lower $S\chi NL$ baryon number densities and chemical potentials.

### Section 2: $SU(3)_L \times SU(3)_R \chi^{PT}$, Baryon & Pseudo-Goldstone Boson Octets, Power Counting, Spherical Representation, Solutions with $\partial_\mu \hat{\pi}_a = 0$

The chiral symmetry of 3 light-quark flavors in QCD, together with symmetry breaking and Goldstone's theorem, makes it possible to obtain an approximate solution to QCD at low energy-momenta and temperatures using $SU3\chi^{PT}$ [1,2,3]. In particular, the non-linear effective Lagrangian has been shown to successfully model the interactions of pions and kaons with baryons, where a perturbation expansion (e.g. in soft momentum $\vec{k}/\Lambda_{\chi SB} \ll 1$, baryon number density $\Psi^\dagger \Psi / f_\pi^2 \Lambda_{\chi SB} \ll 1$ for chiral symmetry breaking scale $\Lambda_{\chi SB} \approx 1 GeV$) has demonstrated predictive power. It has also been shown that all $SU3\chi^{PT}$ analytic quantum loop effects are included in that power counting for very low/soft momentum $\ll \Lambda_{\chi SB}$ [3]. Therefore, $SU3\chi^{PT}$ tree-level calculations are to be regarded as predictions of QCD and the standard model [1,2,3,7]. Warning: among $SU3\chi^{PT}$'s outstanding questions is whether power counting converges with a large strange quark current mass $m_s / \Lambda_{\chi SB} \approx 1/4$ and non-analytic terms [26]. The reader might consult H.Georgi [7] for an introduction and review of $SU3\chi^{PT}$. Useful results and formulae are collected in Appendix 1.

$SU3\chi^{PT}$'s effective-field-theoretic predictive power [1,2,3,7,26] (i.e. its ability to control its analytic quantum loops by power counting, while maintaining a well-ordered perturbation expansion which can be truncated) is to be starkly contrasted with theories of strong interactions which lose their field-theoretic predictive power: e.g. Quark bags and various confinement models of hadronic structure [27]; Strange quark matter and Strange quark stars [28]; Multi-Skyrmions in chiral pseudo-Goldstone symmetry [29]; Models of light and heavy nuclei not demonstrably derivable from $SU(2) \times U(1)\chi^{PT}$ [24]. In contrast, quark-gluon QCD lattice gauge theory calculations [30] control quantum loops and we may hope that the detailed properties of the deuteron may someday be directly calculated there.

It is useful to introduce a "spherical" representation of $SU3\chi^{PT}$ in order to simplify the analysis of chiral liquids

$$\pi_a = f_\pi \hat{\pi}_a \theta; \quad \hat{\pi}_a \hat{\pi}_a = 1; \quad \pi_a F_a = \frac{f_\pi}{2}\theta \hat{\pi}; \tag{10}$$

Because the $SU3_{L+R}$ vector charges simply generate rotations in $\hat{\pi}_a$ space

$$[Q_a^{L+R}, \hat{\pi}_b] = if_{abc}\hat{\pi}_c; \quad [Q_a^{L+R}, \theta] = 0; \tag{11}$$

variation of the pseudo-Goldstone field unit vector $\hat{\pi}_a$ generates the eight $SU3_{L+R}$ vector currents $T_a^\mu$ and CVC equations

$$\partial_\mu T_a^\mu = 0; \quad a = 1,8;$$

$$T_a^\mu = \frac{if_\pi^2}{2} Tr\left([\Sigma^\dagger, F_a]\partial^\mu \Sigma\right) + Tr\overline{B}\gamma^\mu[V_a, B]$$
$$+ DTr\overline{B}\gamma^\mu\gamma^5\{A_a, B\} + FTr\overline{B}\gamma^\mu\gamma^5[A_a, B] \tag{12}$$

$$V_a = \frac{1}{2}(\xi^\dagger F_a \xi + \xi F_a \xi^\dagger); \quad A_a = \frac{1}{2}(\xi^\dagger F_a \xi - \xi F_a \xi^\dagger);$$

In contrast, the eight $SU3_{L-R}$ axial vector currents $T^{5\mu}_a$ and PCAC equations depend on the $\theta$ equation, gotten by variation of the magnitude $\pi_a \pi_a = f_\pi^2 \theta^2$ because

$$\left[Q_a^{L-R}, \theta\right] \neq 0 \tag{13}$$

$$T^{5\mu}_a = \frac{if_\pi^2}{2} Tr\left(\{\Sigma^\dagger, F_a\}\partial^\mu \Sigma\right) + Tr\overline{B}\gamma^\mu[A_a, B]$$
$$+ DTr\overline{B}\gamma^\mu\gamma^5\{V_a, B\} + FTr\overline{B}\gamma^\mu\gamma^5[V_a, B] \tag{14}$$

$$\partial_\mu T^{5\mu}_a \to 0 \quad as \quad L^{SymmetryBreaking}_{\chi PT} \to 0;$$

We identify a specific new class of semi-classical solutions:

$$\partial_\mu \hat{\pi}_a = 0; \tag{15}$$

$$\partial_\mu \Sigma \to i\partial_\mu \theta \hat{\pi} \Sigma; \quad \partial_\mu \xi \to \frac{i}{2}\partial_\mu \theta \hat{\pi} \xi; \tag{16}$$

$$L_{\chi PT} \to \frac{f_\pi^2}{2} \partial_\mu \theta \partial^\mu \theta + Tr\overline{B}(i\gamma^\mu \partial_\mu - m^B)B$$
$$- \frac{1}{2}\partial_\mu \theta \left(DTr\overline{B}\gamma^\mu\gamma^5\{\hat{\pi}, B\} + FTr\overline{B}\gamma^\mu\gamma^5[\hat{\pi}, B]\right) \tag{17}$$
$$+ L^{4-B;Symmetric}_{\chi PT} + L^{SymmetryBreaking}_{\chi PT}$$

Variation of the "SU3 Radius" $\sqrt{\pi_a \pi_a} = f_\pi \theta$ yields an equation for $\theta(t, \vec{x})$:

$$0 = \left(\partial_\mu \frac{\partial}{\partial(\partial_\mu \theta)} - \frac{\partial}{\partial \theta}\right) L_{\chi PT} = f_\pi^2 \partial^2 \theta - \frac{\partial}{\partial \theta} L^{SymmetryBreaking}_{\chi PT}$$
$$- \frac{1}{2}\partial_\mu \left(DTr\overline{B}\gamma^\mu\gamma^5\{\hat{\pi}, B\} + FTr\overline{B}\gamma^\mu\gamma^5[\hat{\pi}, B]\right) \tag{18}$$

### Section 3: $npK^0\overline{K^0}$ Effective Lagrangian, Nucleon Point-Coupling Exchange Terms, Dirac Equation

Nucleons are immersed in a neutral kaon $\theta$ condensate:

$$B \to \begin{bmatrix} 0 & 0 & p \\ 0 & 0 & n \\ 0 & 0 & 0 \end{bmatrix}; \quad \hat{\pi} \to \hat{K} = \begin{bmatrix} 0 & 0 & 0 \\ 0 & 0 & \hat{K}^0 \\ 0 & \hat{\overline{K}^0} & 0 \end{bmatrix}; \quad \hat{\overline{K}^0}\hat{K}^0 = 1; \tag{19}$$

$$\Sigma = 1 + is_\theta \hat{K} + (c_\theta - 1)\hat{K}\hat{K}; \quad \xi = 1 + is_{\theta/2}\hat{K} + (c_{\theta/2} - 1)\hat{K}\hat{K}; \tag{20}$$

with $s_\theta = \sin(\theta)$, $c_\theta = \cos(\theta)$. It follows from Equations (20) and (A4.2) that

$$DTr\overline{B}\gamma^\mu\gamma^5\{\hat{\pi}, B\} + FTr\overline{B}\gamma^\mu\gamma^5[\hat{\pi}, B] = 0$$
$$0 = f_\pi^2 \partial^2\theta - \frac{\partial}{\partial\theta}L_{\chi_{PT}}^{SymmetryBreaking} \tag{21}$$

Including terms to order $\Lambda_{\chi SB}, \Lambda_{\chi SB}^0$ power counting (up to linear in $m_S$), the $SU3\chi^{PT}$ Lagrangian is greatly simplified, especially when up and down quark masses are neglected in the symmetry-breaking sector:

$$L_{\chi_{PT}}^{B;Symmetric} \to L_{\chi_{PT}}^{\Psi;Symmetric} = \overline{\Psi}(i\gamma^\mu\partial_\mu - m^N)\Psi$$

$$L_{\chi_{PT}}^{4-B;Symmetric} \to L_{\chi_{PT}}^{4-\Psi;Symmetric} = -\frac{C_{200}^A}{2f_\pi^2}(\overline{\Psi}\gamma^A\Psi)(\overline{\Psi}\gamma_A\Psi);$$

$$L_{\chi_{PT}}^{\pi;Symmetric} \to L_{\chi_{PT}}^{K;Symmetric} = \frac{f_\pi^2}{2}\partial_\mu\theta\partial^\mu\theta; \tag{22}$$

$$L_{\chi_{PT}}^{\pi;SymmetryBreaking} \to L_{\chi_{PT}}^{K;SymmetryBreaking} = -2f_\pi^2 m_K^2 s_{\theta/2}^2;$$

$$L_{\chi_{PT}}^{B;SymmetryBreaking} \to L_{\chi_{PT}}^{\Psi;SymmetryBreaking} = \overline{\Psi}(-\tilde{m} + m^N)\Psi = 2\sigma_{\pi N}s_{\theta/2}^2\overline{\Psi}\begin{bmatrix}\beta^{pK} & 0\\ 0 & \beta^{nK}\end{bmatrix}\Psi;$$

$$L_{\chi_{PT}}^{SymmetryBreaking} \to L_{\chi_{PT}}^{K;SymmetryBreaking} + L_{\chi_{PT}}^{\Psi;SymmetryBreaking} + L_{\chi_{PT}}^{4-\Psi;SymmetryBreaking}$$

Higher order $\Lambda_{\chi SB}^{-1}$ terms $L_{\chi_{PT}}^{4-\Psi;SymmetryBreaking}$ are considered in Section 6.4. We define

$$\Psi = \begin{bmatrix}p\\n\end{bmatrix}; \quad m^N = 0.939 GeV; \quad M \to diag(0, 0, m_s);$$

$$\tilde{m} = \begin{bmatrix}\tilde{m}^p & 0\\ 0 & \tilde{m}^n\end{bmatrix} = \tilde{m}^+ + 2t_3\tilde{m}^-; \quad \tilde{m}^\pm = \frac{1}{2}(\tilde{m}^p \pm \tilde{m}^n);$$

$$\tilde{m}^p = m^N - 2\beta^{pK}\sigma_{\pi N}s_{\theta/2}^2; \quad \beta^{pK} = (2a_3 + a_2)m_s/\sigma_{\pi N}; \tag{23}$$

$$\tilde{m}^n = m^N - 2\beta^{nK}\sigma_{\pi N}s_{\theta/2}^2; \quad \beta^{nK} = (2a_3 + a_2 + a_1)m_s/\sigma_{\pi N};$$

$$\sigma_{\pi N} = 60 MeV; \quad m_u \to 0; \quad m_d \to 0; \quad m_S \to 250 MeV;$$

$$\gamma^A = 1, \gamma^\mu, i\sigma^{\mu\nu}, i\gamma^\mu\gamma^5, \gamma^5; \quad A = 1 \to 16 = S, V, T, A, P;$$

Since $\beta^{pK}\sigma_{\pi N} > 0$; $\tilde{m}^p < m^N$; and $\beta^{nK}\sigma_{\pi N} > 0$; $\tilde{m}^n < m^N$; the effective nucleon mass (to order $\Lambda_{\chi SB}^0$ up to linear in $m_S$) is smaller [15,16] in the presence of the kaon condensate, as are effective kaon masses in the presence of nucleons (Figure 1)

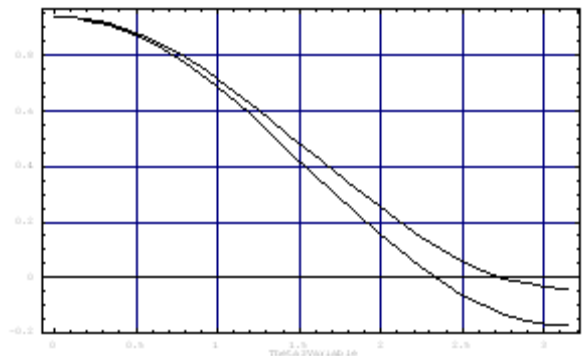

Figure 1: Reduced proton (upper line) and neutron (lower line) masses $\tilde{m} = Diag(\tilde{m}^p, \tilde{m}^n)$ from $L_{\chi PT}^{\Psi;SymmetryBreaking}$ as functions of $\theta$ condensate.

## 3.2 $SU3\chi^{PT}$ **Nucleon Point-Coupling Exchange Terms**

The empirical nuclear models of P.Manakos & T.Mannel [31] were specifically built to include nucleon exchange terms. J.Friar, D. Madland & B.W.Lynn [8] first identified their successor "Nuclear Skyrme Models" [9,12] as derivable from $SU2xU1\chi^{PT}$ chiral liquids. To order $\Lambda_{\chi SB}^0$ the only 4-nucleon point-coupling terms allowed by local $SU3\chi^{PT}$ symmetry [5] are (see Appendix 1)

$$L_{\chi PT}^{4-B;Symmetric} \to L_{\chi PT}^{4-\Psi;Symmetric} = -\frac{C_{200}^A}{2f_\pi^2}(\overline{\Psi}\gamma^A\Psi)(\overline{\Psi}\gamma_A\Psi); \qquad (24)$$

Note that no isospin operators appear:

$$\vec{t} = \frac{1}{2}\vec{\sigma}_{Pauli}; \qquad (25)$$

Quantum loop power counting requires inclusion of nucleon exchange interactions, which are ~ same size as direct interactions, in the analysis of $SU3\chi^{PT}$ states. In order to simplify analysis of the states to Hartree (i.e. rather than Hartree-Fock), we introduce an extended effective Lagrangian $L_{\chi PT}^{4-\Psi;Symmetric,Exchange}$ which explicitly includes exchange contributions to point coupling interactions:

$$L_{\chi PT}^{4-\Psi;Symmetric} \to L_{\chi PT}^{4-\Psi;Symmetric,Exchange}$$
$$= -\frac{C_{200}^A}{2f_\pi^2}(\overline{\Psi}\gamma^A\Psi)(\overline{\Psi}\gamma_A\Psi) \qquad (26)$$
$$+ \frac{\overline{C_{200}^A}}{4f_\pi^2}\left[(\overline{\Psi}\gamma^A\Psi)(\overline{\Psi}\gamma_A\Psi) + (\overline{\Psi}\gamma^A 2t_3\Psi)(\overline{\Psi}\gamma_A 2t_3\Psi)\right]$$

Isospin operators have appeared [19]. When building quantum states, Hartree treatment of $L_{\chi PT}^{4-\Psi;Symmetric,Exchange}$ is equivalent to Hartree-Fock treatment of $L_{\chi PT}^{4-\Psi;Symmetric}$ by Fierz re-arrangement:

$$\begin{bmatrix}\overline{C_{200}^S}\\ \overline{C_{200}^V}\\ \overline{C_{200}^T}\\ \overline{C_{200}^A}\\ \overline{C_{200}^P}\end{bmatrix} = \frac{1}{4}\begin{bmatrix}1 & 4 & 6 & 4 & 1\\ 1 & -2 & 0 & 2 & -1\\ 1 & 0 & -2 & 0 & 1\\ 1 & 2 & 0 & -2 & -1\\ 1 & -4 & 6 & -4 & 1\end{bmatrix}\begin{bmatrix}C_{200}^S\\ C_{200}^V\\ C_{200}^T\\ C_{200}^A\\ C_{200}^P\end{bmatrix}; \qquad (27)$$

The non-relativistic limit of 4-nucleon point-coupling interactions is properly gotten by Fierz re-arranging first and then taking the non-relativistic limit. The result $L_{\chi PT}^{4-\Psi;Symmetric,Exchange} \to$

$$-\frac{(C_{200}^S + C_{200}^V)}{2f_\pi^2}(\Psi^\dagger\Psi)(\Psi^\dagger\Psi) - \frac{(C_{200}^A - 2C_{200}^T)}{2f_\pi^2}(\Psi^\dagger\vec{\sigma}\Psi)\bullet(\Psi^\dagger\Psi^\dagger\vec{\sigma}\Psi\Psi$$

$$+\frac{\left(\overline{C^S_{200}} + \overline{C^V_{200}}\right)}{4f_\pi^2}\left[(\Psi^\dagger\Psi)(\Psi^\dagger\Psi) + (\Psi^\dagger 2t_3\Psi)(\Psi^\dagger 2t_3\Psi)\right]$$
$$+\frac{\left(\overline{C^A_{200}} - 2\overline{C^T_{200}}\right)}{4f_\pi^2}\left[(\Psi^\dagger\vec{\partial}\Psi)\bullet(\Psi^\dagger\vec{\partial}\Psi) + (\Psi^\dagger 2t_3\vec{\partial}\Psi)\bullet(\Psi^\dagger 2t_3\vec{\partial}\Psi)\right] \quad (28)$$

differs significantly from the analogous expression that follow from the non-relativistic 4-fermion Lagrangian [5,6] used in the literature on effective 2-nucleon forces in light nuclei: e.g. it depends on $C^P_{200}$.

### 3.3 Nucleon Dirac Equations

To order $\Lambda^0_{\chi SB}$ up to terms linear in $m_S$

$$\left(i\partial_\mu\gamma^\mu - \Theta\right)\Psi = 0; \quad \overline{\Psi}\left(i\overleftarrow{\partial}_\mu\gamma^\mu - \Theta\right) = 0;$$
$$\Theta = \tilde{m} + \left(\frac{C^A_{200}}{f_\pi^2}(\overline{\Psi}\gamma^A\Psi) - \frac{\overline{C^A_{200}}}{2f_\pi^2}\left[(\overline{\Psi}\gamma^A\Psi) + (\overline{\Psi}\gamma^A 2t_3\Psi)(2t_3)\right]\right)\gamma_A; \quad (29)$$

The pure-nucleon vector currents are conserved (see appendix 3)

$$\partial^\mu J_{a,\mu} = 0; \quad a = 1,2,3; \quad (30)$$
$$\partial^\mu J_{8,\mu} = 0; \quad (31)$$

as are the baryon number current

$$J^\mu_{Baryon} = \frac{2}{\sqrt{3}}J^\mu_8 = \overline{p}\gamma^\mu p + \overline{n}\gamma^\mu n; \quad \partial_\mu J^\mu_{Baryon} = 0; \quad (32)$$

and the electromagnetic current

$$T^\mu_{QED} = T^\mu_3 + \frac{1}{\sqrt{3}}T^\mu_8 = J^\mu_3 + \frac{1}{\sqrt{3}}J^\mu_8 = \overline{p}\gamma^\mu p \quad (33)$$
$$\partial_\mu T^\mu_{QED} = 0;$$

Even though explicit pion and eta fields have been set to zero, their quantum numbers re-appear (from nucleon bi-linears and 4-nucleon terms) in $\Pi^{\overline{\Psi}\Psi}_a, \Pi^{\overline{\Psi}\Psi}_8$ (Appendix 3), which play an important role in ensuring PCAC:

$$\partial^\mu J^5_{a,\mu} = \Pi^{\overline{\Psi}\Psi}_a = 2\tilde{m}^+\left(\overline{\Psi}i\gamma^5 t_a\Psi\right) \quad (34)$$
$$+\left[\delta_{a3}\tilde{m}^- + \frac{2C^P_{200}}{f_\pi^2}\left(\overline{\Psi}t_a\Psi\right)\right]\left(\overline{\Psi}i\gamma^5\Psi\right) + ++; \quad a = 1,2,3$$

$$\partial^\mu J^5_{8,\mu} = \Pi^{\overline{\Psi}\Psi}_8 = \sqrt{3}\tilde{m}^+\left(\overline{\Psi}i\gamma^5\Psi\right) + \left[2\sqrt{3}\tilde{m}^- - \frac{2\sqrt{3}\overline{C^P_{200}}}{f_\pi^2}\left(\overline{\Psi}t_3\Psi\right)\right]\left(\overline{\Psi}i\gamma^5 t_3\Psi\right) + ++;$$

In this paper, we focus on (proton even, neutron even) states. The physical structure and dynamics of (proton odd, neutron odd) states lies beyond the scope of this paper. Among the usual nucleon bi-linear currents (Appendix 3) $J^\mu_\pm, J^{5\mu}_\pm$ mediate transitions between (proton even, neutron even) states and (proton odd, neutron odd) states. Their analysis, and that of $\partial^\mu J^5_{1,\mu}, \partial^\mu J^5_{2,\mu}, \Pi_1^{\overline{\Psi}\Psi}, \Pi_2^{\overline{\Psi}\Psi}$, lies beyond the scope of this paper.

## Section 4: $SU(3)_L \times SU(3)_R \chi^{PT}$ Thomas-Fermi Mean Fields, CVC, PCAC, Chiral Nucleon Liquids, Kaon Condensate

### 4.1 Thomas –Fermi Mean Fields

We are interested here in semi-classical solutions identifiable as quantum chiral nucleon liquids with the following physical properties:
- Ground state;
- Spin=0;
- Spherical (e.g. closed shells);
- Even number of protons, even number of neutrons;

and will use the approximations [13,19]:
- No anti-nucleon sea;
- Thomas-Fermi mean field;
- Spherical liquid drops;
- 4-nucleon point-couplings including exchange terms;
- Relativistic Mean Field Point Coupling Hartree-Fock (RMF-PC-HF)

The Thomas-Fermi mean field liquid approximation is adequate for the purposes of this paper (e.g. density functionals would be over-kill!): details can be found in Appendix 2. It replaces neutrons with expectation values over free neutron spinors with effective mass, 3-momentum and energy $m_*^n, \vec{k}^n, E^n = \sqrt{(\vec{k}^n)^2 + (m_*^n)^2}$ respectively and protons with expectation values over free proton spinors with effective mass, 3-momentum and energy respectively $m_*^p, \vec{k}^p, E^p = \sqrt{(\vec{k}^p)^2 + (m_*^p)^2}$. Nucleon bi-linear forms are replaced by their expectation values in the chiral nucleon liquid:

$$\overline{\Psi}\Psi \to \langle\overline{\Psi}\Psi\rangle; \quad \Psi^\dagger\Psi \to \langle\Psi^\dagger\Psi\rangle;$$
$$\overline{\Psi}\gamma^A\Psi \to \langle\overline{\Psi}\gamma^A\Psi\rangle \to 0; \quad \gamma^A = \gamma^1, \gamma^2, \gamma^3, i\sigma^{\mu\nu}, i\gamma^\mu\gamma^5, \gamma^5; \quad (35)$$

Since we focus on liquid phases, our notation will be simplified (in the remainder of the paper, except Appendices 1, 3, 4) to assume liquid-state expectation values $\langle\ \rangle$ of nucleon bi-linears (Appendix 2), without exchange contributions. Exchange contributions will be displayed explicitly. Then, within the liquid phase:

$$\partial^\mu J_{3,\mu} = 0; \quad \vec{J}_3 = 0;$$
$$\partial^\mu J_{8,\mu} = 0; \quad \vec{J}_8 = 0;$$
$$\Pi_3 = 0; \quad \partial^\mu J^5_{3,\mu} = 0; \quad J^5_{3,\mu} = 0; \quad (36)$$
$$\Pi_8 = 0; \quad \partial^\mu J^5_{8,\mu} = 0; \quad J^5_{8,\mu} = 0;$$

We now show that, in the liquid approximation, a spherically symmetric $SU3\chi^{PT}$ nucleon liquid drop with kaon condensate $\theta(r)$ and

$$f_\pi^2 \partial^2\theta - \frac{\partial}{\partial\theta} L^{SymmetryBreaking}_{\chi_{PT}} = 0; \quad \partial_0\theta = 0; \quad (37)$$

satisfies (as required) all relevant CVC and PCAC equations. The eight $SU3_{L+R}$ vector currents $T_a^\mu$ and eight $SU3_{L-R}$ axial vector currents $T^{5\mu}_a$ are displayed (in general for nucleons and neutral kaons, not just in the liquid) in Appendix 4. Of these, four vector and four axial vector currents (A4.7) mediate transitions between (proton even, neutron even) states and (proton odd, neutron odd) states and lie beyond the scope of this paper. Of the remaining four $SU3_{L+R}$ vector and four $SU3_{L-R}$ axial vector currents (A4.6), those which do not vanish in the Thomas-Fermi liquid are:

$$T_3^\mu = \delta_0^\mu \left[ J_3^0 + \frac{1}{2}\left(-J_3^0 + \sqrt{3}J_8^0\right)s_{\theta/2}^2 \right];$$

$$T_8^\mu = \delta_0^\mu \left[ J_8^0 - \frac{\sqrt{3}}{2}\left(-J_3^0 + \sqrt{3}J_8^0\right)s_{\theta/2}^2 \right];$$

$$T^{5\mu}_6 + iT^{5\mu}_7 = -f_\pi^2 \overline{\hat{K}^0} \partial^\mu \theta + \frac{i}{2} s_\theta \overline{\hat{K}^0} \delta_0^\mu \left[-J_3^0 + \sqrt{3}J_8^0\right];$$

$$T^{5\mu}_6 - iT^{5\mu}_7 = -f_\pi^2 \hat{K}^0 \partial^\mu \theta - \frac{i}{2} s_\theta \hat{K}^0 \delta_0^\mu \left[-J_3^0 + \sqrt{3}J_8^0\right];$$

(38)

which, when $\partial^0 \theta = 0$, satisfy the relevant CVC and PCAC equations:

$$\partial_\mu T_3^\mu = 0; \quad \partial_\mu T_8^\mu = 0;$$

$$\partial_\mu (T_6^\mu \pm iT_7^\mu) = 0;$$

$$\partial_\mu T^{5\mu}_3 = 0; \quad \partial_\mu T^{5\mu}_8 = 0;$$

(39)

$$\partial_\mu (T^{5\mu}_6 + iT^{5\mu}_7) = -f_\pi^2 \overline{\hat{K}^0} \partial^2 \theta = -\overline{\hat{K}^0} \frac{\partial}{\partial \theta} L^{SymmetryBreaking}_{\chi PT};$$

$$\partial_\mu (T^{5\mu}_6 - iT^{5\mu}_7) = -f_\pi^2 \hat{K}^0 \partial^2 \theta = -\hat{K}^0 \frac{\partial}{\partial \theta} L^{SymmetryBreaking}_{\chi PT};$$

Because ordinary chiral heavy nuclear liquids [8,9,12], $S\chi NL$ [13,14] and $SNL$ [13] satisfy all relevant $SU3_{\chi^{PT}}$ CVC and PCAC equations in the liquid phase (and have avoided the dangers listed in the Introduction), they are solutions of the tree level semi-classical liquid equations of motion.

### 4.2 Chiral Nucleon Liquids

Properties of the pure-nucleon liquid (including exchange terms) can be derived from the extended effective nucleon Lagrangian

$$L^{\Psi;Exchange}_{\chi PT} \to L^{\Psi;Exchange}_{\chi PT;Liquid}$$

$$= \overline{\Psi}(i\gamma^\mu \partial_\mu - \tilde{m})\Psi - \frac{C^S_{200}}{2f_\pi^2}(\overline{\Psi}\Psi)(\overline{\Psi}\Psi) - \frac{C^V_{200}}{2f_\pi^2}(\Psi^\dagger \Psi)(\Psi^\dagger \Psi)$$

$$+ \frac{\overline{C^S_{200}}}{4f_\pi^2}\left[(\overline{\Psi}\Psi)(\overline{\Psi}\Psi) + (\overline{\Psi}2t_3\Psi)(\overline{\Psi}2t_3\Psi)\right]$$

$$+ \frac{\overline{C^V_{200}}}{4f_\pi^2}\left[(\Psi^\dagger \Psi)(\Psi^\dagger \Psi) + (\Psi^\dagger 2t_3\Psi)(\Psi^\dagger 2t_3\Psi)\right]$$

(40)

whose Thomas-Fermi mean field details are displayed in Appendix 2. Key to the physics are the reduced effective nucleon masses, to order $\Lambda^0_{\chi SB}$ linear in $m_S$, in the presence of the kaon $\theta$ condensate:

$$C^S_{200} - \frac{1}{2}\overline{C^S_{200}} < 0; \tag{41}$$

$$m^p_* = m^N - 2\beta^{pK}\sigma_{\pi N}s^2_{\theta/2} + \frac{C^S_{200}}{f^2_\pi}\overline{\Psi}\Psi - \frac{\overline{C^S_{200}}}{f^2_\pi}\bar{p}p;$$

$$m^n_* = m^N - 2\beta^{nK}\sigma_{\pi N}s^2_{\theta/2} + \frac{C^S_{200}}{f^2_\pi}\overline{\Psi}\Psi - \frac{\overline{C^S_{200}}}{f^2_\pi}\bar{n}n; \tag{42}$$

The nucleon sector, effective liquid Lagrangian and solution mathematics (for $\theta$) used in this paper (to order $\Lambda^0_{\chi SB}$ linear in $m_S$) are identical to that of B.W.Lynn [19] with the substitutions:

$$\beta^{pK}, \beta^{nK} \to [\beta^{\Psi\pi}]_{BWLynn1993}; \quad m^2_K \to [m^2_\pi]_{BWLynn1993}; \quad \frac{\theta}{2} \to [\theta]_{BWLynn1993}; \tag{43}$$

### 4.3 Kaon $\theta$ Condensate, Roll-around-ology

Collecting useful expressions to order $\Lambda^0_{\chi SB}$ and up to linear $m_S$:

$$L^{\Psi;SymmetryBreaking}_{\chi PT} = 2[\beta^{pK}\sigma_{\pi N}\bar{p}p + \beta^{nK}\sigma_{\pi N}\bar{n}n - f^2_\pi m^2_K]s^2_{\theta/2}$$

$$f^2_\pi\partial^2\theta - 2[\beta^{pK}\sigma_{\pi N}\bar{p}p + \beta^{nK}\sigma_{\pi N}\bar{n}n - f^2_\pi m^2_K]\frac{\partial}{\partial\theta}s^2_{\theta/2} = 0; \tag{44}$$

$$\frac{dP^\Psi}{d\theta} = -\bar{p}p\frac{\partial m^p_*}{\partial\theta} - \bar{n}n\frac{\partial m^n_*}{\partial\theta}; \quad L^{SymmetryBreaking}_{\chi PT} = -U^K = -2f^2_\pi m^2_K s^2_{\theta/2};$$

These are more generally re-written (e.g. to include our order $\Lambda^{-1}_{\chi SB}$ linear $m_S$ terms):

$$0 = f^2_\pi\partial^2\theta - \frac{\partial L^{SymmetryBreaking}_{\chi PT}}{\partial\theta} = f^2_\pi\partial^2\theta - \frac{d}{d\theta}(P^\Psi - U^K);$$

$$L^{SymmetryBreaking}_{\chi PT} = L^{K;SymmetryBreaking}_{\chi PT} + L^{\Psi;SymmetryBreaking}_{\chi PT} + L^{4-\Psi;SymmetryBreaking}_{\chi PT}; \tag{45}$$

which follows directly from the conservation of the energy-momentum tensor of a perfect chiral nucleon liquid [19]. If $(T^K_{\chi PT;Liquid})^{\mu\nu}$ is the usual energy-momentum tensor for the pseudo-scalar field (with $\partial_\mu\hat{\pi}_a = 0$) and $u^\nu$ is the velocity of the liquid

$$\partial_\nu(T_{\chi PT;Liquid})^{\mu\nu} = 0;$$

$$(T^{\Psi;Exchange}_{\chi PT;Liquid})^{\mu\nu} = (E^\Psi + P^\Psi)u^\mu u^\nu - g^{\mu\nu}P^\Psi; \tag{46}$$

$$(T_{\chi PT;Liquid})^{\mu\nu} = (T^K_{\chi PT;Liquid})^{\mu\nu} + (T^{\Psi;Exchange}_{\chi PT;Liquid})^{\mu\nu};$$

The liquid velocity $u^\nu = (1,0,0,0)$ in its rest frame. Static spherical chiral liquid drops

$$\partial_0\theta = 0; \quad \partial_0 E^\Psi = \partial_0 P^\Psi = 0; \tag{47}$$

considered here, $S\chi NL$ and ordinary chiral nuclear liquids (re-scaled radius $\tilde{r}$) obey

$$\left(\frac{d^2}{d\tilde{r}^2} + \frac{2}{\tilde{r}}\frac{d}{d\tilde{r}}\right)\theta = -\frac{dV_{Newton}}{d\theta}; \quad V_{Newton} = \frac{P^\Psi - U^K}{4f^4_\pi}; \quad \tilde{r} = 2f_\pi r; \tag{48}$$

This "Roll-around-ology ODE" (48) can be interpreted to great advantage as Newtonian potential motion with "time $\tilde{r}$", "distance $\theta$", "potential energy $V_{Newton}$" and "friction $\frac{2}{\tilde{r}}\frac{d\theta}{d\tilde{r}}$" diminishing with "time" [18,13,19,14].

## Section 5: Ordinary Chiral Nuclear $SU(2) \times U(1)\chi^{PT}$ Liquid Drops, Nuclear Skyrme Models

To identify the ground state of even-even spin-zero spherical closed-shell heavy nuclei in $SU(3)\chi^{PT}$, we examine a pure-nucleon $\theta = 0$ liquid phase in the Isospin-Hypercharge subgroup (simply called "$SU(2) \times U(1)\chi^{PT}$")

$$[SU(2) \times U(1)]_{L+R} \times [SU(2) \times U(1)]_{L-R} << SU(3)_L \times SU(3)_R; \quad (49)$$

A large pre-existing class of high accuracy "Nuclear Skyrme Models" [9,12,10,11; Appendix 7] were first identified by J.Friar, D.G.Madland & B.W.Lynn [8] as derivable from $SU2 \times U1\chi^{PT}$ liquid ($\Lambda_{\chi SB}^n, n = 1,0,-1,-2$) operators:

1. By concentrating on the ground state of even-even spin-zero spherical closed-shell heavy nuclei, nuclear Skyrme models obey a much simplified $SU2 \times U1\chi^{PT}$ (i.e. the set of liquid operators << total set of operators);
2. Without prior consideration of chiral liquid $SU2 \times U1\chi^{PT}$, B.A.Nikolaus, T.Hoch & D.G.Madland [12] fit 9 coefficients (spanning a huge range $10^{-4} MeV^{-2} \to 10^{-18} MeV^{-8}$) to the properties of 3 heavy nuclei and predicted the properties of another 57 heavy nuclei quite accurately. The observational success of their model (with improvements [9]) is competitive with other nuclear models [24]: Binding energies (to $\pm 0.15\%$ of their values); Charge radius ($\pm 0.2\%$); Diffraction radius ($\pm 0.5\%$); Surface thickness ($\pm 1.5\%$); Spin-orbit splittings ($\pm 5\%$); Pairing gaps (to $\pm 0.05$MeV); Observed Isotonic chains, Fission barriers, etc. to various high accuracies;
3. When their coefficients were rescaled as appropriate to $SU2 \times U1\chi^{PT}$ liquids [8], these obeyed $1/\Lambda_{\chi SB}$ power counting through order $\Lambda_{\chi SB}^{-2}$ with chiral coefficients ~1 (with 2 exceptions since improved [9]). The obedience of nuclear Skyrme model coefficients to $1/\Lambda_{\chi SB}$ power counting [8] in $SU2xU1\chi^{PT}$ is now called "Naturalness" in the heavy nuclear structure literature;
4. Symmetric and asymmetric (finite) nuclear liquid drops (and bulk nuclear matter) in nuclear Skyrme models are therefore (almost) $SU2xU1\chi^{PT}$: they both over-count and are missing certain liquid operators (Appendix 7);
5. We have shown that Nuclear Skyrme models satisfy all relevant $SU3_{L+R} \times SU3_{L-R}$ CVC and PCAC equations in the liquid phase and are solutions of the tree level semi-classical liquid equations of motion. The non-vanishing currents are:

$$T_3^\mu = \delta_0^\mu J_3^0; \quad T_8^\mu = \delta_0^\mu J_8^0; \quad (50)$$

G.Gelmini & B.Ritzi [32] independently identified S.A.Chin & J.D.Walecka's theory of bulk symmetric $N = Z$ nuclear matter (without point-coupling exchange terms)

[20] as derivable from $SU2xU1\chi^{PT}$ to order $\Lambda^0_{\chi SB}$. They did not consider finite or asymmetric $N \neq Z$ chiral nuclear liquid drops.

In this section, we use T.Burvenich, D.G.Madland, J.A.Maruhn & P.-C.Reinhard's [9] Lagrangian to order $\Lambda_{\chi SB}, \Lambda^0_{\chi SB}$: higher order terms are discussed in Appendix 7. For spherical nuclei with nuclear density everywhere flat in radius, and $Z, N, A$ (protons, neutrons and baryon number), the relative baryon number densities are fixed:

$$\theta = 0; \quad p^\dagger p = \frac{Z}{A} \Psi^\dagger \Psi; \quad n^\dagger n = \frac{N}{A} \Psi^\dagger \Psi;$$

$$m_*^p = m^N + \frac{C^S_{200}}{f_\pi^2} \overline{\Psi}\Psi - \frac{\overline{C^S_{200}}}{f_\pi^2} \overline{p}p; \quad m_*^n = m^N + \frac{C^S_{200}}{f_\pi^2} \overline{\Psi}\Psi - \frac{\overline{C^S_{200}}}{f_\pi^2} \overline{n}n; \quad (51)$$

The strategy is to re-write everything as functions of the baryon number chemical potentials $\mu_B^p, \mu_B^p$; and densities $n^\dagger n, p^\dagger p$ and fit $C^S_{200}, C^V_{200}, \overline{C^S_{200}}, \overline{C^V_{200}}$ to nuclear data.

## 5.2 $SU2xU1\chi^{PT}$ Symmetric $Z = N$ Chiral Nuclear Liquid Drops

The prescient results of S.A.Chin & J.D.Walecka [20] are gotten by inserting

$$Z = N; \quad p^\dagger p = n^\dagger n = \frac{1}{2} \Psi^\dagger \Psi; \quad \overline{p}p = \overline{n}n = \frac{1}{2} \overline{\Psi}\Psi; \quad (52)$$

and fitting to the bulk properties of symmetric nuclear matter:

$$[\mu_B]_{NuclearMatter} = m^N - 15.75 MeV; \quad [\Psi^\dagger \Psi]_{NuclearMatter} = 0.001485 GeV^3 = 1.847 f_\pi^3; \quad (53)$$

Examination of $L^{\Psi;Exchange}_{\chi PT;Liquid}$ for $Z = N$ shows it is sensitive only to two combinations

$$\frac{1}{2}[C_S^2]_{SAChin1974} = -\frac{1}{2}\left(C^S_{200} - \frac{1}{2}\overline{C^S_{200}}\right) = 1.313;$$

$$\frac{1}{2}[C_V^2]_{SAChin1974} = \frac{1}{2}\left(C^V_{200} - \frac{1}{2}\overline{C^V_{200}}\right) = 0.963; \quad (54)$$

Using these values of $[C_S^2, C_V^2]_{SAChin1974}$, the nuclear binding energy, baryon number density and asymmetry ($N \neq Z$) energy were fit to the volume and asymmetry energy terms in the ingenious Weizsacker empirical mass formula [24] with good accuracy $a_1 = 15.75 MeV; \quad a_4 = 21.8 MeV$. Bulk nuclear compressibility is too high. Since no nuclear surface terms appear in $L^{\Psi;Exchange}_{\chi PT;Liquid}$, the nuclear surface is too sharp (i.e. a step-function). Since the full set of point-coupling exchange terms are not included, asymmetric $N \neq Z$ (infinite or finite) nuclear matter in this theory violates power counting and is therefore not derivable from $SU2xU1\chi^{PT}$ chiral liquids

## 5.3 $SU2xU1\chi^{PT}$ Asymmetric $N \neq Z$ Chiral Nuclear Liquid Drops

We retain the coefficients $[C_S^2, C_V^2]_{SAChin1974}$ for symmetric heavy nuclei (e.g. $^{40}_{20}Ca_{40}$) as they are constrained by nuclear experiment: e.g. proton electric charge density. Unfortunately, nuclear experimental constraints on the other two combinations of chiral coefficients are weak: e.g. nuclear neutron density is less well determined [42]. We choose the other coefficients to <u>simultaneously</u> allow $S\chi NL$ in Sections 6, 7, 8:

$$\left(-C^S_{200} + \overline{C^S_{200}}\right) = 0.62; \quad \left(C^V_{200} - \overline{C^V_{200}}\right) = 0.24; \tag{55}$$

$$\frac{1}{2}C^S_{200} = -2.32; \quad \frac{1}{2}C^V_{200} = 1.81; \quad \frac{1}{2}\overline{C^S_{200}} = -2.00; \quad \frac{1}{2}\overline{C^V_{200}} = 1.69;$$

Although the example coefficients (55) obey $\Lambda^0_{\chi SB}$ power counting, they may have a fine-tuning problem. We fit the bulk properties of $^{82}_{126}Pb_{208}$ with $(Z-N)/A = -0.212$ (regarded as a chiral nuclear liquid drop) with the chiral coefficients in equation (55)

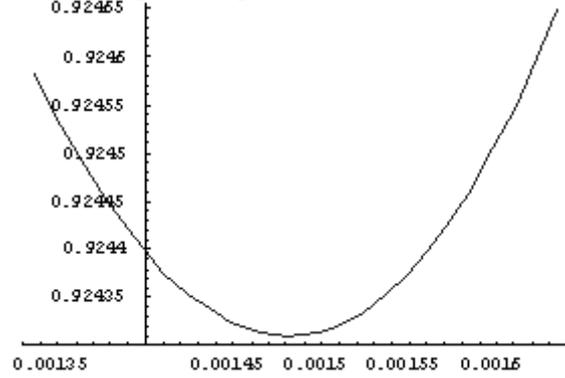

Figure 2: $\frac{E^\Psi}{\Psi^\dagger \Psi}(GeV)$ vs. $\Psi^\dagger \Psi (GeV^3)$ for $^{82}_{126}Pb_{208}$

with a graphical method in Figure 2 [20] which uses the 1$^{st}$ law of thermodynamics

$$P^\Psi = \left(\Psi^\dagger \Psi\right)^2 \frac{\partial}{\partial(\Psi^\dagger \Psi)}\left[\frac{E^\Psi}{\Psi^\dagger \Psi}\right] \tag{56}$$

to identify the lowest point as asymmetric nuclear matter at $P^\Psi = 0$: i.e. neglecting Coulomb forces, stable liquid drops of any radius exist at zero external pressure. The nuclear binding energy, baryon number density and asymmetry ($N \neq Z$) energy still reproduce the volume and asymmetry energy terms in the Weizsacker formula [24]

$$a_1 = 15.75 MeV; \quad a_4 = 23.7 MeV; \tag{57}$$

with good accuracy. Bulk nuclear compressibility remains too high and the nuclear surface too sharp. Yet since a full set of $\sim \Lambda^0_{\chi SB}$ point-coupling exchange terms are included, both symmetric and asymmetric (finite or infinite) nuclear liquid drops obey $SU2 \times U1\chi^{PT}$.

### Section 6: Macroscopic Electrically Neutral $\infty S\chi NL$: Saturating $SU3\chi^{PT}$ Solutions with $\theta \neq 0$

We now make the crucial technical observation of this paper. Due to nucleon point-coupling exchange interactions, the Dirac equation for both microscopic and macroscopic chiral neutron liquids to order $\Lambda^0_{\chi SB}$ up to terms linear in $m_S$

$$\left(i\partial_\mu \gamma^\mu - m^n_* - \frac{1}{f_\pi^2}(C^V_{200} - \overline{C^V_{200}})(n^\dagger n)\gamma^0\right)n = 0; \tag{58}$$

$$m^n_* = m^N + \frac{1}{f_\pi^2}(C^S_{200} - \overline{C^S_{200}})(\bar{n}n) - 2\beta^{nK}\sigma_{\pi N}s^2_{\theta/2}; \quad \beta^{nK} = m_S(2a_3 + a_2 + a_1) \approx 9.28;$$

depends on different combinations (i.e. $C^S_{200} - \overline{C^S_{200}}, C^V_{200} - \overline{C^V_{200}}$) than used for symmetric chiral nuclear liquid drops (Section 5). This allows $S\chi NL$ to (mathematically) <u>co-exist</u> with ordinary heavy nuclei in $SU3\chi^{PT}$ with fixed four-nucleon point couplings (40). (Note: for simplicity we neglect kaon-condensate driven exchange terms here). In contrast, Reference 14 has (in effect) set $\overline{C^S_{200}} = \overline{C^V_{200}} = 0$, thereby neglecting neutron point-coupling exchange terms and violating $SU3\chi^{PT}$ power counting to order $\Lambda^0_{\chi SB}$. Known nuclear-medium effects are perturbative [10] so, baring some miraculous huge medium-dependent renormalization (and the resultant loss of predictive power), $C^2_S, C^2_V$ cannot be chosen (with $C^S_{201}, \overline{C^S_{201}} = 0$) so that <u>both</u> ordinary chiral nuclear liquid drops and $S\chi NL$ emerge from the same theory.

The "Fermion Roll-around-ology" analogy of the $\theta$ condensate equation (48) with Newtonian potential motion [18,19] gives a powerful graphical technique to scan quickly for the existence of Q-Ball non-topological solitons through the very large phase space of parameters (e.g. chiral coefficients, chemical potentials, densities), while using our natural intuition about hills and valleys and the Newtonian motion of a non-relativistic particle "rolling around" through them (i.e. existence & uniqueness proofs, search and construction methods for solutions). Macroscopic (almost infinite) neutral saturating chiral liquid $\infty S\chi NL$ is analogous with (almost infinite) symmetric nuclear matter [18], corresponding to Newtonian potential motion of a particle moving between hills of (almost) equal height, shaped as in the <u>dashed</u> line in Figure 3. We seek solutions at zero external pressure $V_{Newton} \to 0$; $\theta \to 0$; $\tilde{r} \to \infty$; at the top of the left-hand hill. The particle starts near the very top of the right-hand hill with zero initial velocity: (i.e. baryon chemical potential is actually slightly increased over $\infty S\chi NL$ and the hill slightly raised above (the left hand hill) zero since "energy" is dissipated by "friction"). It waits a very long (almost infinite) time there while overcoming friction (almost infinite saturating interior I); drops quickly through the $V_{Newton} < 0$ valley and climbs the left-hand hill (microscopically thin surface S ~ $f_\pi^{-1}$); coming to rest at the top of the left-hand hill (vacuum V).

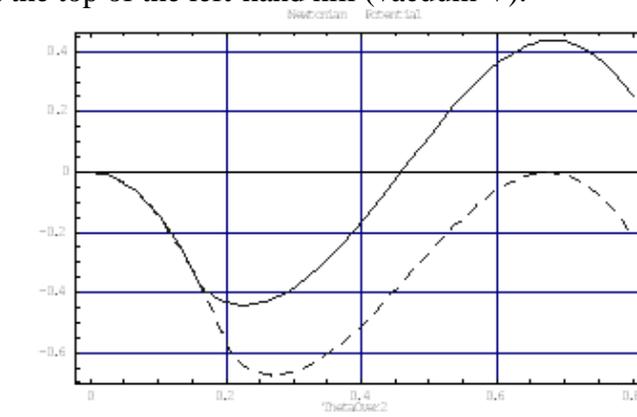

<u>Figure 3</u>: Newtonian Potentials $(C^S_{200} - \overline{C^S_{200}}) = -0.62$, $(C^V_{200} - \overline{C^V_{200}}) = 0.24$ as functions of $\theta$ condensate. Q-Balls can be microscopic ($\mu^n_B = 0.922 GeV$ solid line) or macroscopic ($\mu^n_B = [\mu^n_B]_{\infty S\chi NL} = 0.907 GeV$ dashed line).

It is easy to see graphically that two conditions must be met in order that saturating (almost infinite) macroscopic strange neutral $\infty S\chi NL$ Q-Balls exist:

- "Zero Total Pressure in $\infty S\chi NL$" equation
$$0 = \left[4 f_\pi^4 V_{Newton}\right]_{\infty S\chi NL} = \left[P^n - U^K\right]_{\infty S\chi NL} \quad (59)$$

- "Conservation of Chiral Axial-vector Currents in $\infty S\chi NL$" equation
$$0 = -\left[\partial_\mu (T^{5\mu}_6 \pm i T^{5\mu}_7)\right]_{\infty S\chi NL} = \left[f_\pi^2 \partial^2 \theta\right]_{\infty S\chi NL} = \left[\frac{\partial}{\partial \theta} L^{SymmetryBreaking}_{\chi PT}\right]_{\infty S\chi NL}$$
$$= \left[\frac{d}{d\theta}(P^n - U^K)\right]_{\infty S\chi NL} = \left[4 f_\pi^4 \frac{\partial V_{Newton}}{\partial \theta}\right]_{\infty S\chi NL} ; \quad (60)$$

**6.2 Solutions to order $\Lambda^0_{\chi SB}$ and up to linear $m_S$ in $SU3\chi PT$**

Generally
1. In order to stay (still uncomfortably!) away from the edge of applicability [1,2,3,7] of $SU3\chi^{PT}$, this paper takes $\frac{[n^\dagger n]_{\infty S\chi NL}}{[\Psi^\dagger \Psi]_{NuclearMater}} \leq 6;$ (61)
2. $\infty S\chi NL$ is required to be self-bound (in zero external pressure) against dispersion to free neutrons $[\mu^n_*]_{\infty S\chi NL} < m^N.$ (62)
3. The dashed and dotted lines in Figures $4-7$ are explained in Section 6.3: "Limits on $(C^S_{200} - \overline{C^S_{200}})$ and $(C^V_{200} - \overline{C^V_{200}})$ for given $[n^\dagger n]_{\infty S\chi NL}$";

The Conservation of Chiral Axial-vector Currents in $\infty S\chi NL$ equation (60) shows:
4. Axial chiral currents are conserved:
$$\left[\partial_\mu (T^{5\mu}_6 \pm i T^{5\mu}_7)\right]_{\infty S\chi NL} = \left[2(\beta^{nK} \sigma_{\pi N} \bar{n} n - f_\pi^2 m_K^2) \frac{\partial}{\partial \theta} s^2_{\theta/2}\right]_{\infty S\chi NL} = 0; \quad (63)$$

5. The scalar density is constant (greatly simplifying the algebra)
$$\theta_{\infty S\chi NL} \neq 0; \quad [\bar{n} n]_{\infty S\chi NL} = \frac{f_\pi^2 m_K^2}{\beta^{nK} \sigma_{\pi N}} = \frac{f_\pi^2 m_K^2}{m_S (2a_3 + a_2 + a_1)}; \quad (64)$$

6. $\infty S\chi NL$ tends to high baryon number density, due entirely to experimentally measured parameters in the kaon-nucleon symmetry breaking sector.
$$\frac{[n^\dagger n]_{\infty S\chi NL}}{[\Psi^\dagger \Psi]_{NuclearMater}} > \frac{[\bar{n} n]_{\infty S\chi NL}}{[\Psi^\dagger \Psi]_{NuclearMater}} = \frac{m_K^2}{1.847 f_\pi \beta^{nK} \sigma_{\pi N}} \approx 2.593; \quad (65)$$

7. The one-to-one (1to1) relation between $[m^n_*]_{\infty S\chi NL}$ and $[n^\dagger n]_{\infty S\chi NL}$ is independent of $C^S_{200}, C^V_{200}, \overline{C^S_{200}}, \overline{C^V_{200}}$:
$$[n^\dagger n]_{\infty S\chi NL} = \frac{1}{3\pi^2}\left([k^n_F]_{\infty S\chi NL}\right)^3; \quad [\mu^n_*]_{\infty S\chi NL} = \left[\sqrt{(k^n_F)^2 + (m^n_*)^2}\right]_{\infty S\chi NL};$$
$$\frac{f_\pi^2 m_K^2}{m_S(2a_3 + a_2 + a_1)} = 2[m^n_*]_{\infty S\chi NL} \int_o^{[k^n_F]_{\infty S\chi NL}} \frac{d^3 k^n}{(2\pi)^3}\left((\vec{k}^n)^2 + ([m^n_*]_{\infty S\chi NL})^2\right)^{-\frac{1}{2}}$$
$$= \left[\frac{m^n_*}{2\pi^2}\left(k^n_F \mu^n_* - \frac{1}{2}(m^n_*)^2 \ln\left(\frac{\mu^n_* + k^n_F}{\mu^n_* - k^n_F}\right)\right)\right]_{\infty S\chi NL} ; \quad (66)$$

and is used below to eliminate $[m_*^n]_{\infty S\chi NL}$ in favor of $[n^\dagger n]_{\infty S\chi NL}$ (Appendix 5).

8. $C_{200}^V - \overline{C_{200}^V}$ determines a 1to1 relationship between $[\mu_B^n]_{\infty S\chi NL}$ and $[n^\dagger n]_{\infty S\chi NL}$.

$$[\mu_B^n]_{\infty S\chi NL} = \left[\mu_*^n + \frac{C_{200}^V - \overline{C_{200}^V}}{f_\pi^2} n^\dagger n\right]_{\infty S\chi NL} ; \qquad (67)$$

9. To order $\Lambda_{\chi SB}^0$ up to linear $m_S$ we find high baryon chemical potential and number density in qualitative (but not in numerical) agreement with the conclusions of Reference 15 (Appendix 5);

$$[\mu_B^n]_{\infty S\chi NL} \geq 0.896 GeV; \quad \frac{[n^\dagger n]_{\infty S\chi NL}}{[\Psi^\dagger \Psi]_{NuclearMatter}} \sim 4.84 - 6; \qquad (68)$$

10. $C_{200}^S - \overline{C_{200}^S}$ determines a 1to1 relationship between $[s_{\theta/2}^2]_{\infty S\chi NL}$ and $[n^\dagger n]_{\infty S\chi NL}$

$$[m_*^n]_{\infty S\chi NL} = m^N + \left[\frac{(C_{200}^S - \overline{C_{200}^S})}{f_\pi^2} \bar{n}n - 2\beta^{nK}\sigma_{\pi N} s_{\theta/2}^2\right]_{\infty S\chi NL} ; \qquad (69)$$

Finally, the Zero Total $\infty S\chi NL$ Pressure equation (59) gives another 1to1 relationship between $[\mu_B^n]_{\infty S\chi NL}$ and $[n^\dagger n]_{\infty S\chi NL}$ determined by $(C_{200}^S - \overline{C_{200}^S})$ shown in Figures 4, 5.

$$\left[\left(\frac{1}{2}\mu_B^n - \frac{1}{4}\mu_*^n\right)n^\dagger n - \left(m^N - \frac{3}{4}m_*^n + \frac{(C_{200}^S - \overline{C_{200}^S})}{2f_\pi^2}\bar{n}n\right)\bar{n}n\right]_{\infty S\chi NL} = 0; \qquad (70)$$

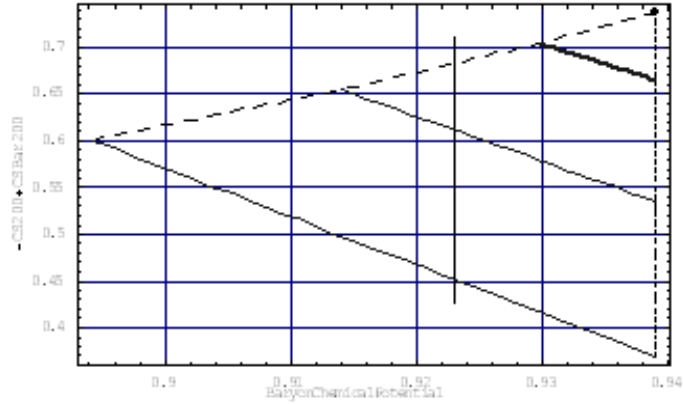

Figure 4: $(-C_{200}^S + \overline{C_{200}^S})$ vs $[\mu_B^n]_{\infty S\chi NL}(GeV)$ to order $\Lambda_{\chi SB}^0$ and linear $m_S$. See text for details. Macroscopic Q-Balls to the left of the vertical solid line have binding energy per nucleon deeper than ordinary nuclear liquids.

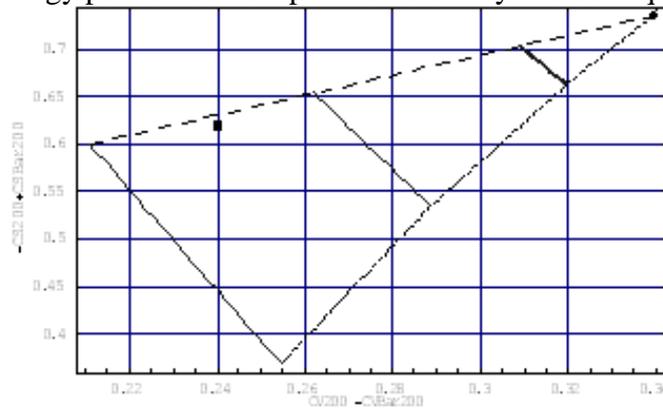

Figure 5: Eliminate $[\mu_B^n]_{\infty S\chi NL}$ to plot $(-C_{200}^S + \overline{C_{200}^S})$ vs $(C_{200}^V - \overline{C_{200}^V})$ to order $\Lambda^0_{\chi SB}$ and linear $m_S$. See text for details.

The slanted solid lines and big dot in Figures 4, 5 have constant baryon number density: 3 solid lines with increasing thickness and big dot have respective densities

$$\frac{[n^\dagger n]_{\infty S\chi NL}}{[\Psi^\dagger \Psi]_{NuclearMatter}} = 5.990, 5.492, 5.086, 4.840; \tag{71}$$

The dotted line has $[\mu_*^n]_{\infty S\chi NL} = m^N$ so the liquid can't disperse energetically to free neutrons inside the dotted line. "Physical" pairs $(C_{200}^S - \overline{C_{200}^S}, C_{200}^V - \overline{C_{200}^V})$ are defined inside the (rough) triangles subtended by dotted, dashed and thinnest solid lines. Because of "friction" $\frac{2}{\tilde{r}}\frac{d\theta}{d\tilde{r}}$, the boundary for the existence of self-bound macroscopic non-topological soliton Q-Balls is actually slightly inside these triangles.

### 6.3 Limits on $(-C_{200}^S + \overline{C_{200}^S})$ and $(C_{200}^V - \overline{C_{200}^V})$ for given $[n^\dagger n]_{\infty S\chi NL}$;

Following [18,13], B.W.Lynn [19] used Roll-around-ology (48) to show (including 4-nucleon point interaction terms) that the mathematics of (microscopic and macroscopic) liquid drops in $SU2 \times U1\chi^{PT}$ maps self-consistently onto Newtonian potential motion, whose well-known existence/uniqueness properties he used to classify non-strange $\chi NL$ according to the existence versus non-existence of Q-Ball non-topological solitons (in the phase space of baryon chemical potential and number density): i.e. Roll-around-ology is also a systematic search method to find those circumstances in which solutions do not exist. That "unphysical" region (un-shaded region of Figure 3 there) was excluded (numerically, not algebraically) because $V_{Newton}$ was "…discontinuous or multi-valued (as a function of $k_F^n$) or the right-hand hill (Figure 2 there, Figure 3 here) was too low". To illustrate the idea, we fix $(C_{200}^S - \overline{C_{200}^S}) = -0.62$; $(C_{200}^V - \overline{C_{200}^V}) = 0.24$; from the ordinary chiral nuclear liquid example of Section 5 and show $V_{Newton}$ (48) for two baryon number chemical potentials $\mu_B^n = 0.922 GeV$; $\mu_B^n = [\mu_B^n]_{\infty S\chi NL} = 0.907 GeV$; (i.e. microscopic and macroscopic $S\chi NL$ respectively). Newtonian potential motion is well-defined because $\frac{ds_{\theta/2}^2}{dk_F^n} \geq 0$ and is continuous everywhere along each $V_{Newton}$. We want "physical" pairs $(C_{200}^S - \overline{C_{200}^S}, C_{200}^V - \overline{C_{200}^V})$ where $\frac{ds_{\theta/2}^2}{dk_F^n} \geq 0$ and is continuous everywhere in $V_{Newton}$ so that non-topological solitons (Q-Balls) exist.

1. We know of no a priori reason fixing the signs of $(-C_{200}^S + \overline{C_{200}^S})$, $(C_{200}^V - \overline{C_{200}^V})$;
2. Upper limits on $(-C_{200}^S + \overline{C_{200}^S})$, lower limits on $(C_{200}^V - \overline{C_{200}^V})$: For a given set of 4-fermion chiral coefficients, a graphical method [14] exists to identify systematically when solutions were multi-valued in $k_F^n$ for the reason that $(-C_{200}^S + \overline{C_{200}^S})$ is too high or $(C_{200}^V - \overline{C_{200}^V})$ is too low for a given $[n^\dagger n]_{\infty S\chi NL}$. The constraint can also be stated $[\mu_B^n]_{\infty \chi SNL} \geq ([\mu_B^n]_{\infty \chi SNL})_{Minimum}$: details of our algebraic version are given in Appendix 5. The locus of such constraints are the dashed lines in Figures $4 \to 7, A5.2, A5.3, A8.2, A8.3$. Examination of Figures $4,6, A5.2, A8.2$ shows that, for a given $[\mu_B^n]_{\infty \chi SNL}$, it is also the line of minimum $\frac{[n^\dagger n]_{\infty S\chi NL}}{[\Psi^\dagger \Psi]_{NuclearMatter}}$ [14].

3. Lower limits on $(-C_{200}^S + \overline{C_{200}^S})$, upper limits on $(C_{200}^V - \overline{C_{200}^V})$:
    - The dotted lines of Figures $4 \to 7, A5.2, A5.3, A8.2, A8.3$ have
    $$m^N > [\mu_B^n]_{\infty \chi SNL}; \tag{72}$$
    - As shown in Appendix 8, for higher density a new <u>upper</u> bound
    $$m^N > ([\mu_B^n]_{\infty \chi SNL})_{Maximum} \geq [\mu_B^n]_{\infty \chi SNL}; \tag{73}$$
    on chemical potential emerges from the requirement that $\frac{ds_{\theta/2}^2}{dk_F^n} \geq 0$ everywhere along the path of Newtonian potential motion;

### 6.4 Higher Order $\Lambda_{\chi SB}^{-1}$ terms up to linear $m_S$ in $SU3\chi PT$

It is beyond the scope of this paper to construct <u>complete minimal</u> (order $\Lambda_{\chi SB}^{-2}$) sets:
- Chiral liquid operators with $\theta = 0$ for nuclear Skyrme models: A systematic program of calibration/ calculation/experiment for detailed properties of the ground state of even-even spin-zero spherical closed-shell heavy nuclei in RMF-PC-HF and ordinary nuclear liquid drops (using that set) is necessary to extract nuclear predictions from $SU2 \times U1\chi^{PT}$ (Appendix 7). The results will place strong constraints on $S\chi NL$, especially $(C_{200}^S - \overline{C_{200}^S}, C_{200}^V - \overline{C_{200}^V})$.
- Chiral liquid $SU3\chi^{PT}$ operators with $\theta \neq 0$ are necessary to calculate the detailed properties of $S\chi NL$ drops co-existent with ordinary nuclear liquids;

It is instructive to modify the $SU3\chi^{PT}$ Lagrangian with (an incomplete set of) higher order terms. For simplicity, we choose explicit chiral symmetry breaking 4-fermion point-coupling order $\Lambda_{\chi SB}^{-1}$ operators which do not contribute to vector or axial vector currents:

$$L_{\chi PT}^{\pi, 4-B; SymmetryBreaking} \to L_{\chi PT}^{K, 4-\Psi; SymmetryBreaking; Exchange}$$

$$= f_\pi^2 \Lambda_{\chi SB}^2 \left( 2 \frac{m_S}{\Lambda_{\chi SB}} s_{\theta/2}^2 \right) \tag{74}$$

$$\times \left( -C_{201}^S \left( \frac{\overline{\Psi}\Psi}{f_\pi^2 \Lambda_{\chi SB}} \right) \left( \frac{\overline{\Psi}\Psi}{f_\pi^2 \Lambda_{\chi SB}} \right) + \frac{\overline{C_{201}^S}}{2} \left[ \left( \frac{\overline{\Psi}\Psi}{f_\pi^2 \Lambda_{\chi SB}} \right) \left( \frac{\overline{\Psi}\Psi}{f_\pi^2 \Lambda_{\chi SB}} \right) + \left( \frac{\overline{\Psi}2t_3\Psi}{f_\pi^2 \Lambda_{\chi SB}} \right) \left( \frac{\overline{\Psi}2t_3\Psi}{f_\pi^2 \Lambda_{\chi SB}} \right) \right] \right);$$

We have dropped $\Lambda_{\chi SB}^{-3}$ terms (because scalar densities are convenient) and included $\overline{C_{201}^S}$ point coupling exchange terms. For convenience and completeness, all formulae in the Appendices (except Appendix 5) have been modified to include this higher order term: details appear in Appendix 8.

1. The equation for Conservation of Chiral Axial-vector Currents (60) in macroscopic $\infty S\chi NL$ non-topological liquid drops now reads:

$$0 = \left[\partial_\mu (T^{5\mu}_6 \pm iT^{5\mu}_7)\right]_{\infty S\chi NL} = -\left[f_\pi^2 \partial^2 \theta\right]_{\infty S\chi NL} = -\left[\frac{\partial}{\partial \theta} L^{SymmetryBreaking}_{\chi_{PT}}\right]_{\infty S\chi NL}$$

$$= -\left[\left(2\beta^{nK}\sigma_{\pi N}\bar{n}n - 2f_\pi^2 m_K^2 - 2\frac{\left(C^S_{201} - \overline{C^S_{201}}\right)}{f_\pi^2}\frac{m_S}{\Lambda_{\chi SB}}(\bar{n}n)^2\right)\frac{\partial}{\partial \theta}s^2_{\theta/2}\right]_{\infty S\chi NL}; \quad (75)$$

2. For $C^S_{201} - \overline{C^S_{201}} < 0$, its solution

$$\frac{[\bar{n}n]_{\infty S\chi NL}}{[\Psi^\dagger \Psi]_{NuclearMatter}} = \frac{(2a_3 + a_2 + a_1)}{2\alpha_{\infty S\chi NL}\left(-C^S_{201} + \overline{C^S_{201}}\right)}\left(-1 + \sqrt{1 + \frac{4\left(-C^S_{201} + \overline{C^S_{201}}\right)}{(2a_3 + a_2 + a_1)^2}}\right); \quad (76)$$

$$\alpha_{\infty S\chi NL} = \frac{m_S [\Psi^\dagger \Psi]_{NuclearMatter}}{f_\pi^2 m_K^2} = 0.1662;$$

yields significant reductions (compare with Equation 65) in the lower bound for $\infty S\chi NL$ scalar density

$$\left.\frac{[\bar{n}n]_{\infty S\chi NL}}{[\Psi^\dagger \Psi]_{NuclearMatter}}\right|_{\substack{(-C^S_{201}+\overline{C^S_{201}})=1.5 \\ a_3=1.5}} = 2.113; \quad (77)$$

3. B.Ritzi & G.Gelmini [14] identified the $C^S_{201}$ term above and found numerically its importance for lowering baryon number densities and chemical potentials. Although our numerical results disagree with theirs (compare top two rows of Tables 1 and 2), this explains their qualitative conclusion;

4. We worry that, since the effect of higher order $m_S^2, m_S/\Lambda_{\chi SB}$ terms on baryon number densities and chemical potentials can be large, $SU3\chi^{PT}$ predictions for drops of $S\chi NL$ may not converge sufficiently quickly in $m_S$, resulting in significant danger to the predictive power of $SU3\chi^{PT}$ [26];

Figures 6 and 7 show an extreme case [14] with slanted lines and big dot of much lower constant baryon number densities: 3 solid lines with increasing thickness and big dot have respective densities (compare with Equation 71)

$$\frac{[n^\dagger n]_{\infty S\chi NL}}{[\Psi^\dagger \Psi]_{NuclearMatter}} = 6.00, 4.00, 3.66, 3.36 \quad (78)$$

Figures 6 shows that such higher order ($\Lambda^{-1}_{\chi SB}$ and linear $m_S$) terms may also lower baryon chemical potentials significantly

$$[\mu^n_B]_{\infty S\chi NL} \geq 0.807 GeV; \quad (79)$$

when comparing to Equation (68). Details and more order $\Lambda^{-1}_{\chi SB}$ and linear $m_S$ results appear in Appendix 8.

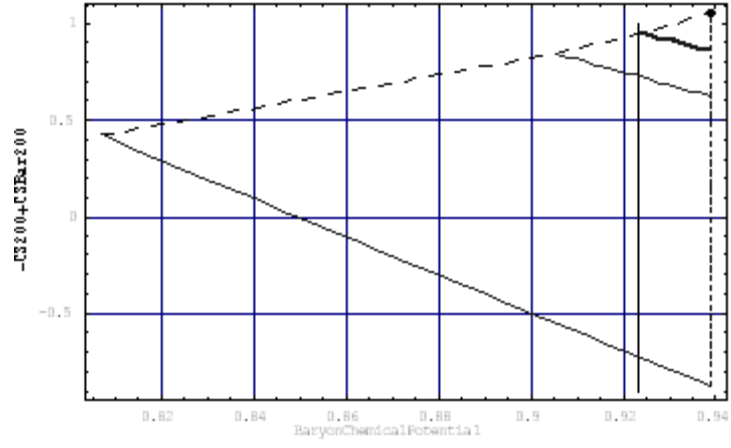

Figure 6: $(-C_{200}^S + \overline{C_{200}^S})$ vs $[\mu_B^n]_{\infty S\chi NL}$ (GeV) to order $\Lambda_{\chi SB}^{-1}$ and linear $m_S$ with $C_{201}^S - \overline{C_{201}^S} = -1.5$ and $a_3 = 1.5$. See text for details. Macroscopic Q-Balls to the left of the vertical solid line have binding energy per nucleon deeper than ordinary nuclear liquids.

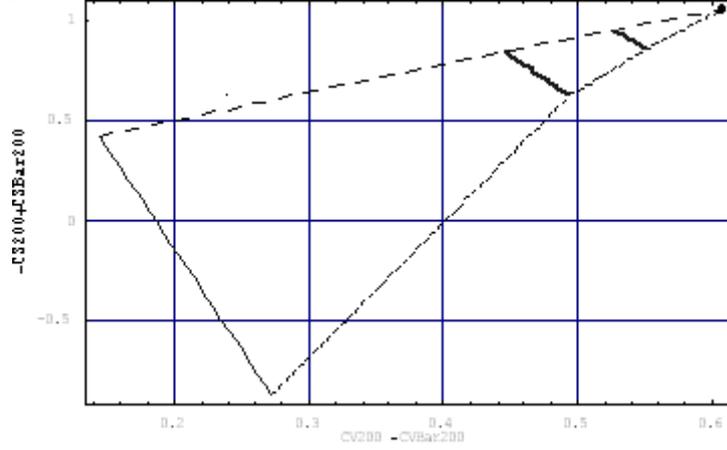

Figure 7: Eliminate $[\mu_B^n]_{\infty S\chi NL}$ to plot $(-C_{200}^S + \overline{C_{200}^S})$ vs $(C_{200}^V - \overline{C_{200}^V})$ to order $\Lambda_{\chi SB}^{-1}$ and linear $m_S$ with $C_{201}^S - \overline{C_{201}^S} = -1.5$ and $a_3 = 1.5$. See text for details.

### 6.5 Numerical Results

Our main numerical results are in Figures 4-7, A5.1-3, A8.1-3 and Table 1

| Table 1 Compare with | $[\mu_B^n]_{\infty \chi SNL}$ (GeV) | $[n^\dagger n]_{\infty S\chi NL}$ $[\Psi^\dagger \Psi]_{Nuclear\ Matter}$ | $[m_*^n]_{\infty \chi SNL}$ (GeV) | $[s_{\theta/2}^2]_{\infty \chi SNL}$ | $-\overline{C_{200}^S}$ $+C_{200}^S$ | $-\overline{C_{200}^V}$ $+C_{200}^V$ | $-\overline{C_{201}^S}$ $+C_{201}^S$ | $a_3$ |
|---|---|---|---|---|---|---|---|---|
| Table2,Row1 Figures 5,8 Section 5,7 | 0.907 | 5.72 | 0.229 | 0.390 | -0.620 | 0.240 | 0.0 | 1.3 |
| Table2,Row2 $S\chi NLStar\,2$ | 0.900 | 4.73 | 0.210 | 0.295 | -0.692 | 0.333 | -1.5 | 1.3 |
|  | 0.930 | 5.09 | 0.258 | 0.329 | -0.705 | 0.309 | 0.0 | 1.3 |
| $S\chi NLStar\,3$ | 0.914 | 5.49 | 0.238 | 0.368 | -0.655 | 0.262 | 0.0 | 1.3 |
|  | 0.918 | 4.33 | 0.230 | 0.261 | -0.788 | 0.401 | -1.5 | 1.3 |

| | | 0.883 | 5.11 | 0.195 | 0.322 | -0.612 | 0.278 | -1.5 | 1.3 |
| | | 0.924 | 3.66 | 0.235 | 0.225 | -0.951 | 0.527 | -1.5 | 1.5 |
| $S\chi NLStar\,1$ | | 0.905 | 4.00 | 0.214 | 0.257 | -0.844 | 0.445 | -1.5 | 1.5 |
| Figures 7,9 Section 7 | | 0.870 | 5.00 | 0.173 | 0.354 | -0.445 | 0.282 | -1.5 | 1.5 |
| $S\chi NLStar\,4$ | | 0.807 | 6.00 | 0.148 | 0.373 | -0.421 | 0.144 | -1.5 | 1.5 |

The numerical results of Reference 14 are quoted in Table 2 with the correspondence

$$C_{200}^S \Rightarrow -[C_S^2]_{B.Ritz\ddot{\imath}1997}; \quad C_{200}^V \Rightarrow [C_V^2]_{B.Ritz\ddot{\imath}1997}; \quad \Lambda_{\chi SB} \Rightarrow \frac{m_K^2}{m_S}; \quad (80)$$

$$C_{201}^S \Rightarrow -\frac{1}{2}[I_S - I_V]_{B.Ritz\ddot{\imath}1997}; \quad \overline{C_{200}^V}, \overline{C_{200}^S}, \overline{C_{201}^S} \Rightarrow [0]_{B.Ritz\ddot{\imath}1997};$$

| **Table 2 [14]** Compare with | $[\mu_B^n]_{\infty\chi SNL}$ (GeV) | $[n^\dagger n]_{\infty S\chi NL}$ $[\Psi^\dagger \Psi]_{Nuclear\,Matter}$ | $[m_*^n]_{\chi SNL}$ (GeV) | $[s_{\theta/2}^2]_{\infty\chi SNL}$ | $-\overline{C_{200}^S} + C_{200}^S$ | $-\overline{C_{200}^V} + C_{200}^V$ | $-\overline{C_{201}^S} + C_{201}^S$ | $a_3$ |
|---|---|---|---|---|---|---|---|---|
| Table1,Row1 | 0.900 | 6.6 | 0.22 | 0.415 | -0.62 | 0.24 | 0.0 | 1.3 |
| Table1,Row2 | 0.900 | 5.3 | 0.21 | 0.319 | -0.72 | 0.36 | -1.5 | 1.3 |

Comparison of the top two rows of Tables 1 and 2 shows significant disagreement in baryon number density, chemical potential and 4-nucleon point-coupling chiral coefficients: e.g. as shown by the small black rectangle in Figure 5, the point $[C_S^2, C_V^2]_{B.Ritz\ddot{\imath}1997} = (0.62, 0.24)$ does not lie (contrary to Reference [14]) on the line of minimum baryon number density. This does not appear to be a typographical error: they mistakenly substitute $\mu_*^n \to \mu_B^n$ in their expressions (see their equations 4.16, 4.17 and contrast with Appendix 2, Equation A2.5) for both scalar density $\bar{n}n$ and pressure $P^{\int n}$ thereby, in principle, mis-balancing $C_{200}^V - \overline{C_{200}^V}$ repulsion, destroying the relativistic covariance of their Thomas-Fermi mean field neutron liquid, violating $T^{5\mu}_6 \pm iT^{5\mu}_7$ PCAC, failing to conserve chiral axial-vector currents in the interior of $\infty S\chi NL$ and destroying Roll-around-ology.

## Section 7: Finite Microscopic Electrically Neutral $S\chi NL$ Droplets

For finite baryon number solitons, $V_{Newton}$ (48) is shaped like the underlined solid line in Figure 3. As usual, we seek solutions at zero external pressure $V_{Newton} \to 0; \quad \theta \to 0; \quad \tilde{r} \to \infty$; at the top of the left-hand hill. The particle starts near the top of the right-hand hill where $m^N > \mu_B^n > [\mu_B^n]_{\infty\chi SNL}$ (well above zero since "energy" is dissipated by "friction") with zero initial velocity; waits there awhile overcoming friction (saturating interior I); quickly drops through the valley and climbs the left-hand hill (thin surface S); coming to rest at the top of the left-hand hill (vacuum V). The strategy is to write $(n^\dagger n, \mu_*^n, m_*^n, \bar{n}n, k_F^n)$ in terms of $(\mu_B^n, \theta)$ and use the Roll-around-ology ODE (48) to calculate $\theta(\mu_B^n, \tilde{r})$ numerically. Figures 8 and 9 show Q-Ball liquid drop $s_{\theta/2}^2$ (thin lower solid line), baryon number density $n^\dagger n$ (dashed line),

$$\mathrm{E}^{Total} = \mathrm{E}^{\Psi} + \frac{f_\pi^2}{2}(\vec{\nabla}\theta)^2 + 2f_\pi^2 m_K^2 s_{\theta/2}^2; \tag{81}$$

which includes the energy density in the soliton surface (dotted line) and finally $-T^0_{Strangeness}$ (thick upper solid line). The last three are normalized by $7[\Psi^\dagger\Psi]_{NuclearMatter}$.

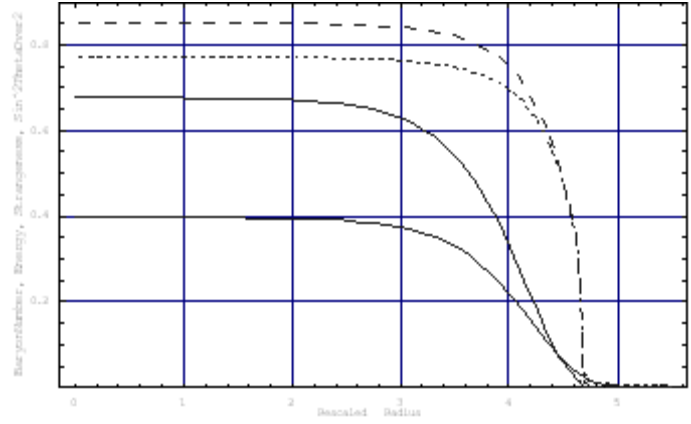

Figure 8: Microscopic Q-Ball to order $\Lambda^0_{\chi SB}$ and linear $m_S$. Baryon number A=502, strangeness S= 285, integrated energy per baryon E/A=0.929 GeV. $(C^S_{200} - \overline{C^S_{200}}) = -0.62$, $(C^V_{200} - \overline{C^V_{200}}) = 0.24$, $C^S_{201} - \overline{C^S_{201}} = 0$, $a_3 = 1.3$. See text.

"Friction" $\frac{2}{\tilde{r}}\frac{d\theta}{d\tilde{r}}$ has played an important role, cutting down significantly the space of chiral coefficients containing finite Q-Balls inside the triangles of Figures 4-7, A5.2-3, A7.2-3. Surface energy and friction will be discussed in detail elsewhere.

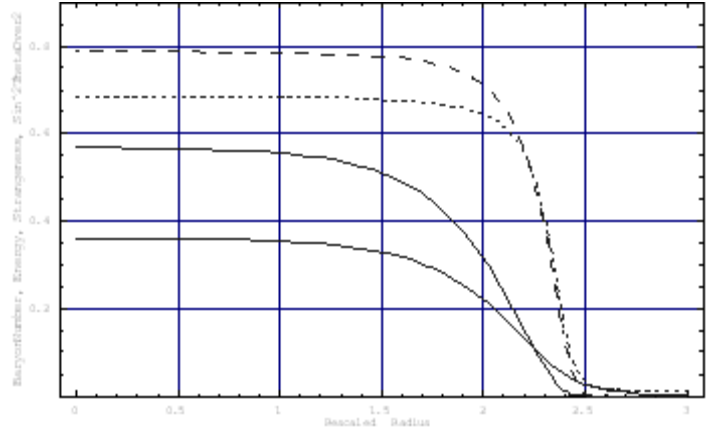

Figure 9: Microscopic non-topological soliton to order $\Lambda^{-1}_{\chi SB}$ and linear $m_S$. A=62, S= 31, integrated energy per baryon E/A=0.919 GeV. See text. $(C^S_{200} - \overline{C^S_{200}}) = -0.445$, $(C^V_{200} - \overline{C^V_{200}}) = 0.282$, $C^S_{201} - \overline{C^S_{201}} = -1.5$, $a_3 = 1.5$

Since we have shown that $S\chi NL$ and ordinary chiral heavy nuclear liquids can co-exist in $SU3\chi PT$, legitimate comparison between them becomes possible. The Weizsacher mass formula, Equation 3.1 of Reference 24 (replacing $a_1 \rightarrow 15.75 MeV$ and $a_4 \rightarrow 23.7 MeV$) gives integrated energy per baryon E/A=0.930 GeV for the most deeply bound stable nucleus, $^{56}_{26}Fe_{30}$. The non-topological solitons in Figures 8 and 9 have E/A=0.929 GeV and E/A=0.919 GeV respectively. The resultant logical possibilities are startling: for certain sets of experimentally allowed chiral coefficients, $S\chi NL$ may be the ground state of a finite collection of nucleons! Ordinary heavy nuclei may only be meta-stable: e.g. the $S\chi NL$ drop of Figure 9 may better approximate the ground state of 62 nucleons.

## Section 8: $S\chi NLStar$ Neutron Stars

A proper treatment [33,18] of $\sim M_{Sun}$ fermion soliton stars (i.e. coupling classical General Relativity (GR) to $SU3\chi^{PT}$, while replacing the nucleons by a liquid in the Thomas-Fermi approximation and an extended and more complex Q-Star non-topological soliton Roll-around-ology) is beyond the scope of this paper and will appear elsewhere [25]. But because $\infty S\chi NL$ densities are so high, we expect gravitational effects to be small in the region where $SU3\chi^{PT}$ is still applicable, so we can make a (very) rough estimate of $S\chi NLStar$ properties by introducing an approximate Equation of State (EOS), treating bulk $S\chi NL$ matter as a perfect fluid and solving the Oppenheimer Volkoff equations [34]. One approximate $S\chi NLStar$ EOS (~massless nucleons) was studied in [13]. Another approximate EOS [25]

$$E - 3P - 4U_0 + \frac{\left(C^V_{200} - \overline{C^V_{200}}\right)}{f_\pi^2 (\mu_B^n)^2}(E+P)^2 = 0; \quad U_0 = \left[U^\Psi + U^K\right]_{\infty S\chi NL}; \quad (82)$$

assumes conservation of chiral axial-vector currents in the stellar interior, varying baryon number density, energy density $E$ and pressure $P$, while (the reader is warned!) artificially (and self-inconsistently) holding the scalar density and condensate angle fixed at the values $\left[\bar{n}n\right]_{\infty S\chi NL}, \left[\theta\right]_{\infty S\chi NL}$. The stellar interior $P(R) \geq 0; \quad R \leq R_{S\chi NLStar}$ with the stellar radius $R_{S\chi NLStar}$ defines $P(R_{S\chi NLStar}) = 0$.

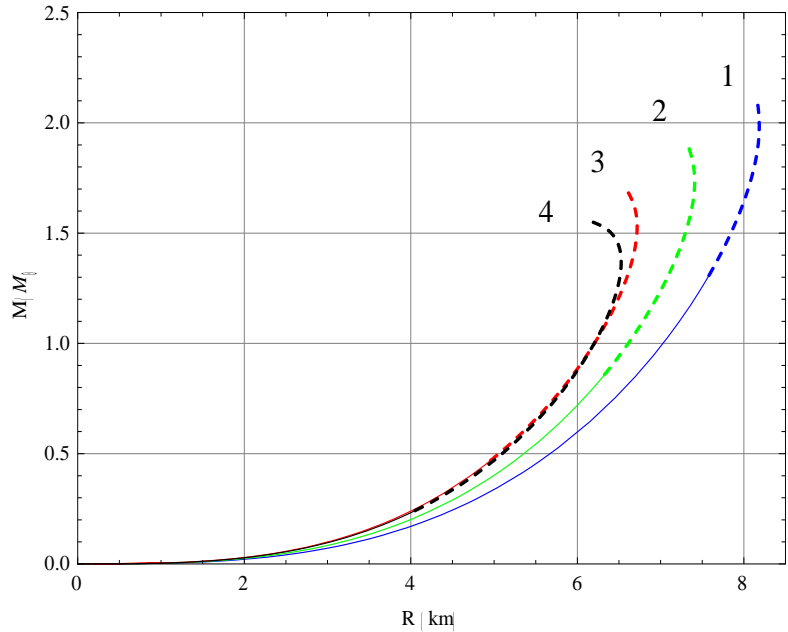

Figure 10: Typical Mass vs Radius curlicue Q-Star profiles. Details of each corresponding macroscopic $\infty S\chi NL$ Q-Ball are given in Table 1.

For small $R_{S\chi NLStar}$, the mass grows as $R^3_{S\chi NLStar}$: $\infty S\chi NL$ matter saturates. For larger central energy density, gravity compresses the stellar interior. Therefore, the $S\chi NLStar$ Mass-Radius relationship follows the distinctive curlicue shape introduced for Q-stars [18,19] and strange quark stars [28]. For central baryon number densities $\frac{[n^\dagger n]_{S\chi NLStar}}{[\Psi^\dagger \Psi]_{NuclearMatter}} > 7$; the $SU3\chi^{PT}$ effective Lagrangian description of hadrons as fundamental particles no longer applies: the fermion Q-Star curlicues are solid lines below this point, dotted (and un-reliable) above it. In the fermion Q-Ball region where gravity is unimportant, mass and baryon number $M, B \sim R^3$, the parameters of the stars are labelled in order of increasing baryon number density and given in Table 1. All Q-Stars shown have binding energy per nucleon deeper than ordinary nuclear liquids.

## Section 9: Going Forward

The biggest question is whether $SU3\chi^{PT}$ is reliable and convergent for such large strange quark masses and high densities. For example we have (at our peril) ignored non-analytic loop contributions: when involving strange quarks, these pose significant dangers for $SU3\chi^{PT}$ relations constraining baryon masses (Gell-Mann Okubo), hyperon S-wave non-leptonic decay, hyperon magnetic moments and baryon axial couplings [26]. Still, we expect interesting theoretical properties of $SBL$ chiral liquids to emerge. Like strange quark matter [28], $S\chi NL$ cannot be dangerously produced at the LHC [13,35]. From the perspective of condensed matter and quantum liquids, nucleon BCS pairing in the ground state of ordinary even-even spin-zero spherical closed-shell heavy nuclei in $SU2xU1\chi^{PT}$ nuclear liquid drops [24,9] will also occur in $S\chi NL$. Parity doubling in a pion S-wave $\theta$ condensate [19] may have analogy in the kaon S-wave $\theta$ condensate. We might hope that a connection with multi-kaon nuclei [36] can be made. The phenomenology (e.g. microscopically thin neutron star Q-Ball surface, pulsar rotation, supernovae remnants, heavy ion collisions, etc) and structure (e.g. inclusion of protons, electron and additional baryon species constituents, etc.) of $S\chi NL$ and $SBL$ is discussed in Reference [13].

A.E. Nelson opened and began analysis [37] of the possibility that a kaon condensate phase (with baryons) might follow the confining/chiral symmetry breaking QCD phase transition in the early universe (e.g. light baryon octet, generation of entropy, minimization of free energy, etc.). Following LNT's suggestion that dark matter might be $SBL$ [13], she began analysis (e.g. baryon evaporation, limits from big bang nucleo-synthesis (BBN), etc.) of "…the slim (logical) possibility that most of the baryon number of the universe could have survived as lumps of SBM (i.e. $SBL$) with baryon number $B \sim 10^{45}$, size $R \sim 100 cm$, mass $M \sim 10^{21} gm$…(and) are candidates for … (cold baryonic) dark matter… It is therefore not necessary to invent new particles to explain dark matter, which is simply made of a new form of baryonic matter… This is also the most natural explanation for the similar mass abundances of ordinary and dark matter. Unfortunately…(with) fluxes $\sim 10^{-38}/cm^2/\sec$, (such) dark matter is impossible to detect [37]." She estimates that this scenario requires binding energy per nucleon $> 185 MeV$. Although achievable in $SBL$ [13], such binding is much deeper than found here or in Reference [14] for $SU3\chi^{PT}$. It is not impossible, however, that such deep $S\chi BL$ binding energies might follow from further analysis (e.g. inclusion of additional baryon species to relieve Fermi pressure, using the large set of unanalyzed order $\Lambda_{\chi SB}^{-1}, \Lambda_{\chi SB}^{-2}$ chiral liquid coefficients contributing to both $S\chi BL$ and ordinary chiral nuclear liquids, etc.). In that case, $S\chi BL$ cold baryonic dark matter may have trapped nucleons sufficiently (in the hadronic phase of the early universe) to evade limits from BBN and the Cosmic Microwave Background.

If anything about $S\chi BL$ turns out to be true (i.e. experimentally observed in Nature), the "Standard $SU3 \times SU2 \times U1 \times$ Semi-classical General Relativity Model" (i.e. the most powerful, accurate, predictive, successful and experimentally verified scientific theory known to humans) would demonstrate that it is still capable of surprising us!

### Acknowledgements

I am deeply grateful to Jennifer A. Thomas and Jonathan R. Ellis. Without their support and encouragement I would not have returned to physics and this paper would not exist. I thank the CERN Theoretical Physics Division and the Department of Physics at University College London for support. I thank B.J.Coffey for many valuable discussions. This paper is dedicated to my teacher Gerald Feinberg (1933-1992) who could, with high precision, tease the periodic table itself (along with positron-ium and muon-ium) out of the standard quantum-field-theoretic model.

# Appendix 1: $SU(3)_L \times SU(3)_R \, \chi^{PT}$ of Baryon & Pseudo-Goldstone Boson Octets

The chiral symmetry of three light quark flavors in QCD, together with symmetry breaking and Goldstone's theorem, makes it possible to obtain an approximate solution to QCD at low energies using an $SU(3)_L \times SU(3)_R$ effective field theory where the degrees of freedom are hadrons [1,2,3,7]. In particular, the non-linear $SU3\chi^{PT}$ effective Lagrangian has been shown to successfully model the interactions of pions and kaons with baryons, where a perturbation expansion (e.g. in soft momentum $\vec{k}/\Lambda_{\chi SB} \ll 1$, baryon number density $\Psi^\dagger \Psi / f_\pi^2 \Lambda_{\chi SB} \ll 1$ for chiral symmetry breaking scale $\Lambda_{\chi SB} \approx 1 GeV$) has demonstrated predictive power. Power-counting in $1/\Lambda_{\chi SB}$ includes all analytic quantum loop effects into experimentally-measured coefficients of $SU(3)_L \times SU(3)_R$ current algebraic operators obedient to the global symmetries of mass-less QCD [3]. Inclusion of operators which obey the symmetries of QCD with light-quark masses generates additional explicit chiral symmetry breaking terms. Therefore, $SU3\chi^{PT}$ tree-level calculations with a power-counting effective Lagrangian are to be regarded as predictions of QCD and the standard model. The reader is warned: among $SU3\chi^{PT}$'s biggest problems are power counting convergence problems [26] caused by the large strange quark mass $m_s/\Lambda_{\chi SB} \approx 1/4$ in (23). A collection of useful formulae [7,38] might include:

1. Unitary 3x3 Matrices $\lambda_a$, asymmetric & symmetric structure constants $f_{abc}$ & $d_{abc}$ and algebra of SU3 generator-charges:

$$
\begin{aligned}
&F_a = \lambda_a / 2; \quad a = 1,8; \quad Tr(F_a F_b) = \delta_{ab}/2; \\
&[F_a, F_b] = if_{abc} F_c; \quad \{F_a, F_b\} = \delta_{ab}/3 + d_{abc} F_c; \\
&[Q_a^{L+R}, Q_b^{L+R}] = if_{abc} Q_c^{L+R}; \quad [Q_a^{L-R}, Q_b^{L-R}] = if_{abc} Q_c^{L+R}; \quad [Q_a^{L+R}, Q_b^{L-R}] = if_{abc} Q_c^{L-R};
\end{aligned}
\quad (A1.1)
$$

2. Pseudo-Goldstone and Baryon Octets;

$$
\pi_a F_a = \frac{1}{\sqrt{2}} \begin{bmatrix} \pi^0/\sqrt{2} + \eta/\sqrt{6} & \pi^+ & K^+ \\ \pi^- & -\pi^0/\sqrt{2} + \eta/\sqrt{6} & K^0 \\ K^- & \overline{K^0} & -2\eta/\sqrt{6} \end{bmatrix} \quad (A1.2)
$$

$$
B = \sqrt{2} B_a F_a = \begin{bmatrix} \Sigma^0/\sqrt{2} + \Lambda/\sqrt{6} & \Sigma^+ & p \\ \Sigma^- & -\Sigma^0/\sqrt{2} + \Lambda/\sqrt{6} & n \\ \Xi^- & \Xi^0 & -2\Lambda/\sqrt{6} \end{bmatrix}
$$

3. In the name of pedagogical simplicity, representations of higher mass are neglected, even though the baryon decuplet (especially $\Delta_{1232}$) is known to carry important nuclear structure [4] and scattering [39] effects;
4. Since $SU3\chi^{PT}$ matrix elements are independent of representation [2], we choose a simple representation [3,7] where, in the symmetric sector, the pseudo-Goldstone octet has <u>only</u> derivative couplings; Then baryon transformations and their coupling to pseudo-Goldstones are <u>local</u>:

$$\Sigma = \exp(2i\pi_a F_a / f_\pi); \quad \Sigma \to \Sigma' = L\Sigma R^\dagger$$

$$\xi = \Sigma^{1/2} = \exp(i\pi_a F_a / f_\pi); \quad \xi \to \xi' = \exp(i\pi'_a F_a / f_\pi);$$

$$\xi' = L\xi U^\dagger = U\xi R^\dagger$$

$$UnitaryGlobal \quad L = \exp(il_a F_a); \quad R = \exp(ir_a F_a);$$

$$UnitaryLocal \quad U(L, R, \pi_a(t,x))$$

$$V_\mu = \frac{1}{2}(\xi^\dagger \partial_\mu \xi + \xi \partial_\mu \xi^\dagger); \quad V_\mu \to UV_\mu U^\dagger + U\partial_\mu U^\dagger;$$

$$A_\mu = \frac{i}{2}(\xi^\dagger \partial_\mu \xi - \xi \partial_\mu \xi^\dagger); \quad A_\mu \to UA_\mu U^\dagger;$$

$$B \to B' = UBU^\dagger;$$

$$D_\mu B = \partial_\mu B + [V_\mu, B]; \quad D_\mu B \to U(D_\mu B)U^\dagger;$$

(A1.3)

5. Include all <u>analytic</u> quantum loop effects for soft momentum $<<$ 1GeV [3]:

$$L_{\chi PT} = -\sum_{l,m,n} C_{lmn} f_\pi^2 \Lambda_{\chi SB}^2 \left(\frac{\partial_\mu}{\Lambda_{\chi SB}}\right)^m \left(\frac{\overline{B}}{f_\pi \sqrt{\Lambda_{\chi SB}}}\right)^l \left(\frac{B}{f_\pi \sqrt{\Lambda_{\chi SB}}}\right)^l \left(\frac{m_{quark}}{\Lambda_{\chi SB}}\right)^n function_{lmn}(\frac{\pi_a}{f_\pi});$$

$$L_{\chi PT} \sim \Lambda_{\chi SB} + (\Lambda_{\chi SB})^0 + \frac{1}{\Lambda_{\chi SB}} + \left(\frac{1}{\Lambda_{\chi SB}}\right)^2 + + + \quad (A1.4)$$

$$l + m + n - 1 \geq 0; \quad C_{lmn;} \sim 1;$$

6. Having approximated $m_u = m_d = 0$, we take self-consistent $\Lambda_{\chi SB} = \frac{m_K^2}{m_S}$.

7. Re-order non-relativistic perturbation expansion in $\partial_0$ to converge with large baryon mass $m^B > \Lambda_{\chi SB}; \quad m^N \approx \Lambda_{\chi SB};$ [5];

8. Tree level calculations only: $SU3\chi^{PT} \to$ strong interaction predictions which are calculable in practice!

9. Include terms to order $\Lambda_{\chi SB}, \Lambda_{\chi SB}^0$ power counting (here we keep up to linear in $m_S$); Separate into "Symmetric" piece (i.e. spontaneous $SU(3)_{L-R}$ breaking with mass-less Goldstones) and a "Symmetry Breaking" piece (i.e. explicit $SU(3)_{L-R}$ breaking, traceable to quark masses) generating eight massive pseudo-Goldstones;

$$L_{\chi PT} = L_{\chi PT}^{Symmetric} + L_{\chi PT}^{SymmetryBreaking}$$

$$L_{\chi PT}^{Symmetric} = L_{\chi PT}^{\pi;Symmetric} + L_{\chi PT}^{B;Symmetric} + L_{\chi PT}^{4-B;Symmetric}; \quad (A1.5)$$

$$L_{\chi PT}^{SymmetryBreaking} = L_{\chi PT}^{\pi;SymmetryBreaking} + L_{\chi PT}^{B;SymmetryBreaking} + L_{\chi PT}^{4-B;SymmetryBreaking};$$

$$L_{\chi PT}^{B;Symmetric} = Tr\overline{B}(i\gamma^\mu D_\mu - m^B)B$$
$$+ DTr\overline{B}\gamma^\mu\gamma^5\{A_\mu, B\} + FTr\overline{B}\gamma^\mu\gamma^5[A_\mu, B];$$

$$L^{4-B;Symmetric}_{\chi PT} \sim \frac{1}{f_\pi^2}(Tr\overline{B}\gamma^A B)(Tr\overline{B}\gamma_A B),\ \frac{1}{f_\pi^2}Tr(\overline{B}\gamma^A B\overline{B}\gamma_A B),\ \frac{1}{f_\pi^2}Tr(\overline{B}\overline{B}BB+++);$$

$$L^{B;SymmetryBreaking}_{\chi PT} = -a_1 Tr\overline{B}(\xi M\xi + h.c.)B - a_2 Tr\overline{B}B(\xi M\xi + h.c.)$$
$$-a_3 Tr\overline{B}B Tr(M(\Sigma-1) + h.c.);$$

$$L^{\pi;Symmetric}_{\chi PT} = \frac{f_\pi^2}{4}Tr\partial_\mu \Sigma \partial^\mu \Sigma^\dagger \qquad (A1.6)$$

$$L^{\pi;SymmetryBreaking}_{\chi PT} = \frac{f_\pi^2}{2}\Lambda_{\chi SB} Tr(M(\Sigma-1) + h.c.)$$

where $\overline{B}\overline{B}, \overline{B}B, etc.$ indicate the order of SU3 index contraction. The experimentally measured chiral coefficients are

$$D = 0.81;\ F = 0.44;\ a_1 = 0.28;\ a_2 = -0.56;\ a_3 = 1.3 \pm 0.2;$$
$$m_u = 0.006 GeV;\ m_d = 0.012 GeV;\ m_S = 0.24 GeV;\ m^B = 1.209 GeV;$$
$$M = Diag(m_u, m_d, m_S);\ \sigma^{\mu\nu} = \frac{1}{2}[\gamma^\mu, \gamma^\nu]; \qquad (A1.7)$$
$$\gamma^A = 1, \gamma^\mu, i\sigma^{\mu\nu}, i\gamma^\mu\gamma^5, \gamma^5;\ A = 1,16 = S,V,T,A,P;$$

The only higher order $\Lambda^{-1}_{\chi SB}$ term $L^{4-B;SymmetryBreaking}_{\chi PT}$ contributes to the explicit kaon-nucleon symmetry breaking sector, is $\theta$-dependent and is considered in Section 6.4.

**Appendix 2: Thomas-Fermi Mean Field Liquid Approximation**

We are interested here in semi-classical nucleon liquids with physical properties:
- Ground state;
- Spin=0;
- Spherical (e.g. closed shells);
- Even number of protons, even number of neutrons;
- No anti-nucleon sea;
- Spherical liquid drops;

The objects of our calculations are, for a liquid drop in its center of mass with radial variable r, the following sets:
- $\mu_B^n, m_*^n(r), k_F^n(r)$ for neutrons: respectively baryon number chemical potential, reduced mass as a function of radius and Fermi-momentum as a function of radius;
- $\mu_B^p, m_*^p(r), k_F^p(r)$ for protons: respectively baryon number chemical potential, reduced mass as a function of radius and Fermi-momentum as a function of radius;
- $\theta(r)$ for the liquid: $SU(3)_{L+R}$ singlet kaon condensate as a function of radius;

The Thomas-Fermi <u>liquid</u> approximation [18,19] replaces neutrons with expectation values over free neutron spinors (with effective reduced mass, 3-momentum and energy $m_*^n, \vec{k}^n, E^n = \sqrt{(\vec{k}^n)^2 + (m_*^n)^2}$) with spin zero and spherical symmetry:

$$\overline{n}n \rightarrow \langle \overline{n}n \rangle \rightarrow \left\langle \frac{m_*^n}{E^n} \right\rangle;$$

$$\bar{n}\gamma^0 n \to \langle \bar{n}\gamma^0 n \rangle \to \langle 1 \rangle; \quad \bar{n}\vec{\gamma}n \to \langle \bar{n}\vec{\gamma}n \rangle \to \left\langle \frac{\vec{k}^n}{E^n} \right\rangle \to 0;$$

$$\bar{n}\sigma^{0j} n \to \langle \bar{n}\sigma^{0j} n \rangle \to \left\langle \frac{(\vec{\sigma}^n \times \vec{k}^n)_j}{E^n} \right\rangle \to 0;$$

$$\bar{n}\sigma^{ij} n \to \langle \bar{n}\sigma^{ij} n \rangle \to \varepsilon_{ijl} \left\langle \left( \vec{\sigma}^n - \frac{\vec{k}^n (\vec{\sigma}^n \cdot \vec{k}^n)}{E^n (E^n + m_*^n)} \right)_l \right\rangle \to 0; \quad (A2.1)$$

$$\bar{n}\gamma^0\gamma^5 n \to \langle \bar{n}\gamma^0\gamma^5 n \rangle \to \left\langle \frac{\vec{\sigma}^n \cdot \vec{k}^n}{E^n} \right\rangle \to 0;$$

$$\bar{n}\vec{\gamma}\gamma^5 n \to \langle \bar{n}\vec{\gamma}\gamma^5 n \rangle \to \left\langle \frac{m_*^n}{E^n}\vec{\sigma}^n + \frac{\vec{k}^n (\vec{\sigma}^n \cdot \vec{k}^n)}{E^n (E^n + m_*^n)} \right\rangle \to 0;$$

$$\bar{n}\gamma^5 n \to \langle \bar{n}\gamma^5 n \rangle = 0;$$

It also replaces protons with expectation values over free proton spinors (with effective reduced mass, 3-momentum and energy $m_*^p, \vec{k}^p, E^p = \sqrt{(\vec{k}^p)^2 + (m_*^p)^2}$) with spin zero and spherical symmetry, with results gotten from those of neutrons by the simple substitution $n \to p$. To simplify our notation, we drop the $\langle\ \rangle$ in the remainder of this Appendix. Within the liquid drop, the total <u>nucleon</u> energy density, pressure density, baryon number density and scalar density $E^\Psi, P^\Psi, \Psi^\dagger\Psi, \bar{\Psi}\Psi$ respectively

$$E^\Psi = E^p + E^n; \quad P^\Psi = P^p + P^n;$$
$$\Psi^\dagger\Psi = p^\dagger p + n^\dagger n; \quad \bar{\Psi}\Psi = \bar{p}p + \bar{n}n; \quad (A2.2)$$

are constructed from

$$\left(T^{\Psi;Exchange}_{\chi PT;Liquid}\right)^{\mu\nu} = \frac{\partial L^{\Psi;Exchange}_{\chi PT;Liquid}}{\partial(\partial_\mu \Psi)}\partial^\nu\Psi - g^{\mu\nu} L^{\Psi;Exchange}_{\chi PT;Liquid};$$

$$E^\Psi = \left(T^{\Psi;Exchange}_{\chi PT;Liquid}\right)^{00}; \quad P^\Psi = \frac{1}{3}\left(T^{\Psi;Exchange}_{\chi PT;Liquid}\right)^{jj}; \quad (A2.3)$$

and liquid properties for neutrons

$$m_*^n = \tilde{m}^n + \frac{C^S_{200}}{f_\pi^2}\bar{\Psi}\Psi - \frac{\overline{C^S_{200}}}{f_\pi^2}\bar{n}n + 4\left(\frac{m_S}{\Lambda_{\chi SB}}s^2_{\theta/2}\right)\left(\frac{C^S_{201}}{f_\pi^2}(\bar{\Psi}\Psi) - \frac{\overline{C^S_{201}}}{f_\pi^2}(\bar{n}n)\right);$$

$$\mu_B^n = \mu_*^n + \frac{C^V_{200}}{f_\pi^2}\Psi^\dagger\Psi - \frac{\overline{C^V_{200}}}{f_\pi^2}n^\dagger n; \quad (A2.4)$$

$$\mu_*^n = \sqrt{(k_F^n)^2 + (m_*^n)^2}; \quad n^\dagger n = 2\int_o^{k_F^n}\frac{d^3k^n}{(2\pi)^3} = \frac{(k_F^n)^3}{3\pi^2};$$

$$\bar{n}n = 2\int_o^{k_F^n}\frac{d^3k^n}{(2\pi)^3}\frac{m_*^n}{\sqrt{(\vec{k}^n)^2 + (m_*^n)^2}} = \frac{m_*^n}{2\pi^2}\left(k_F^n\mu_*^n - \frac{1}{2}(m_*^n)^2\ln\left(\frac{\mu_*^n + k_F^n}{\mu_*^n - k_F^n}\right)\right)$$

$$P^{\int n} = 2\int_o^{k_F^n} \frac{d^3k^n}{(2\pi)^3} \frac{(\vec{k}^n)^2}{3\sqrt{(\vec{k}^n)^2 + (m_*^n)^2}} = \frac{1}{4}\mu_*^n n^\dagger n - \frac{1}{4}m_*^n \overline{n}n$$

$$= \frac{1}{4\pi^2}\left[\mu_*^p\left(\frac{(k_F^n)^3}{3} - \frac{k_F^n(m_*^n)^2}{2}\right) + \frac{1}{4}(m_*^n)^4 \ln\left(\frac{\mu_*^n + k_F^n}{\mu_*^n - k_F^n}\right)\right]; \quad (A2.5)$$

$$E^{\int n} = 2\int_o^{k_F^n} \frac{d^3k^n}{(2\pi)^3}\sqrt{(\vec{k}^n)^2 + (m_*^n)^2} = \frac{3}{4}\mu_*^n n^\dagger n + \frac{1}{4}m_*^n \overline{n}n;$$

$$E^{\int n} - 3P^{\int n} = m_*^n \overline{n}n;$$

The liquid properties of protons are gotten from those of neutrons by the simple substitutions $n \to p, p \to n, \Psi^\dagger\Psi \to \Psi^\dagger\Psi$. The total nucleon energy $E^\Psi$ is written

$$E^\Psi = E^{\int p} + E^{\int n} + \frac{1}{2}\left(\frac{C_{200}^V}{f_\pi^2}(\Psi^\dagger\Psi)^2 - \frac{\overline{C_{200}^V}}{f_\pi^2}\left[(p^\dagger p)^2 + (n^\dagger n)^2\right]\right) \quad (A2.6)$$

$$-\frac{1}{2}\left(\frac{C_{200}^S}{f_\pi^2}(\overline{\Psi}\Psi)^2 - \frac{\overline{C_{200}^S}}{f_\pi^2}\left[(\overline{p}p)^2 + (\overline{n}n)^2\right]\right) - 2\left(\frac{m_S}{\Lambda_{\chi SB}}s_{\theta/2}^2\right)\left(\frac{C_{201}^S}{f_\pi^2}(\overline{\Psi}\Psi)^2 - \frac{\overline{C_{201}^S}}{f_\pi^2}\left[(\overline{p}p)^2 + (\overline{n}n)^2\right]\right)$$

$$= \frac{3}{4}\left(\mu_B^p p^\dagger p + \mu_B^n n^\dagger n\right) - \frac{1}{4}\left(\frac{C_{200}^V}{f_\pi^2}(\Psi^\dagger\Psi)^2 - \frac{\overline{C_{200}^V}}{f_\pi^2}\left[(p^\dagger p)^2 + (n^\dagger n)^2\right]\right) + U^\Psi;$$

$$U^\Psi = \frac{1}{4}\left(\tilde{m}^p \overline{p}p + \tilde{m}^n \overline{n}n\right) - \frac{1}{4}\left(\frac{C_{200}^S}{f_\pi^2}(\overline{\Psi}\Psi)^2 - \frac{\overline{C_{200}^S}}{f_\pi^2}\left[(\overline{p}p)^2 + (\overline{n}n)^2\right]\right) \quad (A2.7)$$

$$-\left(\frac{m_S}{\Lambda_{\chi SB}}s_{\theta/2}^2\right)\left(\frac{C_{201}^S}{f_\pi^2}(\overline{\Psi}\Psi)^2 - \frac{\overline{C_{201}^S}}{f_\pi^2}\left[(\overline{p}p)^2 + (\overline{n}n)^2\right]\right);$$

while the nucleon pressure $P^\Psi$

$$P^\Psi = P^{\int p} + P^{\int n} + \frac{1}{2}\left(\frac{C_{200}^V}{f_\pi^2}(\Psi^\dagger\Psi)^2 - \frac{\overline{C_{200}^V}}{f_\pi^2}\left[(p^\dagger p)^2 + (n^\dagger n)^2\right]\right) \quad (A2.8)$$

$$+\frac{1}{2}\left(\frac{C_{200}^S}{f_\pi^2}(\overline{\Psi}\Psi)^2 - \frac{\overline{C_{200}^S}}{f_\pi^2}\left[(\overline{p}p)^2 + (\overline{n}n)^2\right]\right) + 2\left(\frac{m_S}{\Lambda_{\chi SB}}s_{\theta/2}^2\right)\left(\frac{C_{201}^S}{f_\pi^2}(\overline{\Psi}\Psi)^2 - \frac{\overline{C_{201}^S}}{f_\pi^2}\left[(\overline{p}p)^2 + (\overline{n}n)^2\right]\right)$$

$$= \frac{1}{4}\left(\mu_B^p p^\dagger p + \mu_B^n n^\dagger n\right) + \frac{1}{4}\left(\frac{C_{200}^V}{f_\pi^2}(\Psi^\dagger\Psi)^2 - \frac{\overline{C_{200}^V}}{f_\pi^2}\left[(p^\dagger p)^2 + (n^\dagger n)^2\right]\right) - U^\Psi;$$

is related to the nucleon energy density by the baryon number densities

$$E^\Psi + P^\Psi = \mu_B^p p^\dagger p + \mu_B^n n^\dagger n; \quad (A2.9)$$

A little algebra gives

$$\frac{dP^\Psi}{d\theta} = \left(\frac{ds_{\theta/2}^2}{d\theta}\right) \quad (A2.10)$$

$$\times\left[2\beta^{nK}\sigma_{\pi N}\overline{n}n + 2\beta^{pK}\sigma_{\pi N}\overline{p}p - 2\frac{m_S}{\Lambda_{\chi SB}}\left(\frac{C_{201}^S}{f_\pi^2}(\overline{\Psi}\Psi)^2 - \frac{\overline{C_{201}^S}}{f_\pi^2}\left[(\overline{p}p)^2 + (\overline{n}n)^2\right]\right)\right]$$

Equation A2.10 follows directly from conservation of the energy-momentum tensor for perfect relativistic fluids [19] (i.e. still true when $C_{201}^S, \overline{C_{201}^S} \neq 0$)

$$\frac{dP^\Psi}{d\theta} = \frac{\partial}{\partial\theta}\left(L_{\chi_{PT}}^{\Psi;SymmetryBreaking} + L_{\chi_{PT}}^{4-\Psi;SymmetryBreaking}\right); \quad (A2.11)$$

and is the basis for Q-Ball Roll-around-ology [18,19].

## Appendix 3: Nucleon Bi-linears and $\Pi_a^{\overline{\Psi}\Psi}, \Pi_8^{\overline{\Psi}\Psi}$  $a = 1,2,3$

Including our order $\Lambda_{\chi SB}^{-1}$ terms linear in $m_S$ the Dirac equation (26) becomes

$$(i\partial_\mu \gamma^\mu - \hat{\Theta})\Psi = 0; \quad \overline{\Psi}(i\partial_\mu \gamma^\mu - \hat{\Theta}) = 0; \quad (A3.1)$$

$$\hat{\Theta} = \Theta + 4\left(\frac{m_S}{\Lambda_{\chi SB}} s_{\theta/2}^2\right)\left(\frac{C_{201}^S}{f_\pi^2}(\overline{\Psi}\Psi) - \frac{\overline{C_{201}^S}}{2f_\pi^2}\left[(\overline{\Psi}\Psi) + (\overline{\Psi}2t_3\Psi)(2t_3)\right]\right);$$

so, forming the usual nucleon bi-linears,

$$J_k^\mu = \overline{\Psi}\gamma^\mu t_k \Psi; \quad k = 1,2,3$$

$$J_\pm^\mu = J_1^\mu \pm iJ_2^\mu = \begin{Bmatrix}\overline{p}\gamma^\mu n \\ \overline{n}\gamma^\mu p\end{Bmatrix}; \quad (A3.2)$$

$$J_3^\mu = \frac{1}{2}(\overline{p}\gamma^\mu p - \overline{n}\gamma^\mu n), \quad J_8^\mu = \frac{\sqrt{3}}{2}(\overline{p}\gamma^\mu p + \overline{n}\gamma^\mu n);$$

$$J_k^{5\mu} = \overline{\Psi}\gamma^\mu \gamma^5 t_k \Psi; \quad k = 1,2,3$$

$$J_\pm^{5\mu} = J_1^{5\mu} \pm iJ_2^{5\mu} = \begin{Bmatrix}\overline{p}\gamma^\mu \gamma^5 n \\ \overline{n}\gamma^\mu \gamma^5 p\end{Bmatrix};$$

$$J_3^{5\mu} = \frac{1}{2}(\overline{p}\gamma^\mu \gamma^5 p - \overline{n}\gamma^\mu \gamma^5 n), \quad J_8^{5\mu} = \frac{\sqrt{3}}{2}(\overline{p}\gamma^\mu \gamma^5 p + \overline{n}\gamma^\mu \gamma^5 n)$$

the pure-nucleon pieces of axial vector currents are partially conserved:

$$\partial^\mu J_{a,\mu}^5 = \overline{\Psi}\{i\gamma^5 t_a, \hat{\Theta}\}\Psi = \Pi_a^{\overline{\Psi}\Psi}; \quad a = 1,2,3$$

$$\partial^\mu J_{8,\mu}^5 = \frac{\sqrt{3}}{2}\overline{\Psi}\{i\gamma^5, \hat{\Theta}\}\Psi = \Pi_8^{\overline{\Psi}\Psi}; \quad (A3.3)$$

$$\frac{f_\pi^2}{2}\Pi_a^{\overline{\Psi}\Psi} = f_\pi^2 \tilde{m}^+ (\overline{\Psi}i\gamma^5 t_a \Psi) + \frac{f_\pi^2}{2}\delta_{a3}\tilde{m}^-(\overline{\Psi}i\gamma^5 \Psi)$$

$$+ \left(\hat{C}_{200,201}^S - \frac{1}{2}\overline{\hat{C}_{200,201}^S} - \frac{1}{2}\delta_{a3}\overline{C_{200}^P}\right)(\overline{\Psi}\Psi)(\overline{\Psi}i\gamma^5 t_a \Psi)$$

$$+ \left(C_{200}^P - \frac{1}{2}\overline{C_{200}^P} - \frac{1}{2}\delta_{a3}\overline{\hat{C}_{200,201}^S}\right)(\overline{\Psi}i\gamma^5 \Psi)(\overline{\Psi}t_a \Psi) \quad (A3.4)$$

$$+ \left(C_{200}^T - \frac{1}{2}\overline{C_{200}^T} - \frac{1}{2}\delta_{a3}\overline{C_{200}^T}\right)\varepsilon^{\mu\nu\lambda\sigma}(\overline{\Psi}i\sigma_{\mu\nu}t_a \Psi)(\overline{\Psi}i\sigma_{\lambda\sigma}\Psi)$$

$$+ \varepsilon_{3ab}\left[\overline{C_{200}^V}J_{3,\mu}J_b^{5,\mu} - \overline{C_{200}^A}J_{b,\mu}J_3^{5,\mu}\right];$$

$$\frac{f_\pi^2}{\sqrt{3}} \Pi_8^{\overline{\Psi}\Psi} = f_\pi^2 \tilde{m}^+ \left(\overline{\Psi} i\gamma^5 \Psi\right) + 2 f_\pi^2 \tilde{m}^- \left(\overline{\Psi} i\gamma^5 t_3 \Psi\right)$$

$$+ \left(\hat{C}_{200,201}^S + C_{200}^P - \frac{1}{2}\overline{\hat{C}_{200,201}^S} - \frac{1}{2}\overline{C_{200}^P}\right)\left(\overline{\Psi}\Psi\right)\left(\overline{\Psi} i\gamma^5 \Psi\right)$$

$$- 2\left(\overline{\hat{C}_{200,201}^S} + \overline{C_{200}^P}\right)\left(\overline{\Psi} t_3 \Psi\right)\left(\overline{\Psi} i\gamma^5 t_3 \Psi\right)$$

$$+ \left(C_{200}^T - \frac{1}{2}\overline{C_{200}^T}\right)\varepsilon^{\mu\nu\lambda\sigma}\left(\overline{\Psi} i\sigma_{\mu\nu}\Psi\right)\left(\overline{\Psi} i\sigma_{\lambda\sigma}\Psi\right) \quad (A3.5)$$

$$- 2\overline{C_{200}^T}\varepsilon^{\mu\nu\lambda\sigma}\left(\overline{\Psi} i\sigma_{\mu\nu} t_3 \Psi\right)\left(\overline{\Psi} i\sigma_{\lambda\sigma} t_3 \Psi\right);$$

$$\hat{C}_{200,201}^S = C_{200}^S + 4 C_{201}^S \frac{m_S}{\Lambda_{\chi SB}} s_{\theta/2}^2; \quad \overline{\hat{C}_{200,201}^S} = \overline{C_{200}^S} + 4\overline{C_{201}^S}\frac{m_S}{\Lambda_{\chi SB}} s_{\theta/2}^2; \quad (A3.6)$$

## Appendix 4: Spherical $SU3\chi^{PT}$ representation, Solutions with $\partial_\mu \hat{\pi}_a = 0$, Vector and axial-vector currents relevant to $S\chi NL$ & Ordinary chiral nuclear Liquids

Introducing the spherical $SU3\chi^{PT}$ representation for pseudo-Goldstones

$$\pi_a = f_\pi \hat{\pi}_a \theta; \quad \pi_a F_a = \frac{f_\pi}{2}\theta\hat{\pi}; \quad [\hat{\pi},\Sigma] = [\hat{\pi},\xi] = 0; \quad (A4.1)$$

$$\hat{\pi}_a \hat{\pi}_a = \hat{\pi}^0 \hat{\pi}^0 + \hat{\pi}^+ \hat{\pi}^- + \hat{K}^0 \overset{\wedge}{\overline{K^0}} + \hat{K}^+ \hat{K}^- + \hat{\eta}\hat{\eta} = 1;$$

$$\hat{\pi} = \begin{bmatrix} \hat{\pi}^0 + \hat{\eta}/\sqrt{3} & \hat{\pi}^+ & \hat{K}^+ \\ \hat{\pi}^- & -\hat{\pi}^0 + \hat{\eta}/\sqrt{3} & \hat{K}^0 \\ \hat{K}^- & \overset{\wedge}{\overline{K^0}} & -2\hat{\eta}/\sqrt{3} \end{bmatrix} \quad (A4.2)$$

we search for semi-classical solutions which point in a fixed SU3 field direction

$$\partial_\mu \hat{\pi}_a = 0; \quad (A4.3)$$

with resultant vast simplification:

$$\partial_\mu \Sigma \to i\partial_\mu \theta \hat{\pi} \Sigma; \quad \partial_\mu \Sigma^\dagger \to -i\partial_\mu \theta \hat{\pi} \Sigma^\dagger;$$

$$\partial_\mu \xi \to \frac{i}{2}\partial_\mu \theta \hat{\pi} \xi; \quad \partial_\mu \xi^\dagger \to -\frac{i}{2}\partial_\mu \theta \hat{\pi} \xi^\dagger; \quad (A4.4)$$

$$V_\mu \to 0; \quad A_\mu \to -\frac{1}{2}\partial_\mu \theta \hat{\pi};$$

$$L_{\chi PT} \to \frac{f_\pi^2}{2}\partial_\mu \theta \partial^\mu \theta + Tr\overline{B}(i\gamma^\mu \partial_\mu - m^B)B$$

$$- \frac{1}{2}\partial_\mu \theta \left(D Tr\overline{B}\gamma^\mu \gamma^5 \{\hat{\pi},B\} + F Tr\overline{B}\gamma^\mu \gamma^5 [\hat{\pi},B]\right) \quad (A4.5)$$

$$+ L_{\chi PT}^{4-B;Symmetric} + L_{\chi PT}^{SymmetryBreaking}$$

Variation of the "SU3 Radius" $\sqrt{\pi_a \pi_a} = f_\pi \theta$ yields for $\partial_\mu \hat{\pi}_a = 0$

$$0 = \left(\partial_\mu \frac{\partial}{\partial(\partial_\mu \theta)} - \frac{\partial}{\partial \theta}\right) L_{\chi PT} \tag{A4.6}$$

$$= f_\pi^2 \partial^2 \theta - \frac{\partial}{\partial \theta} L_{\chi PT}^{SymmetryBreaking} - \frac{1}{2} \partial_\mu \left(DTr\bar{B}\gamma^\mu \gamma^5 \{\hat{\pi}, B\} + FTr\bar{B}\gamma^\mu \gamma^5 [\hat{\pi}, B]\right),$$

The relevant four $SU3_{L+R}$ vector and four $SU3_{L-R}$ axial vector currents with protons, neutrons, neutral kaons and $\partial_\mu \hat{\pi}_a = 0$, but not yet evaluated in the liquid, are:

$$T_3^\mu = J_3^\mu + \frac{1}{2}\left(-J_3^\mu + \sqrt{3}J_8^\mu\right)s_{\theta/2}^2;$$

$$T_8^\mu = J_8^\mu - \frac{\sqrt{3}}{2}\left(-J_3^\mu + \sqrt{3}J_8^\mu\right)s_{\theta/2}^2;$$

$$T_6^\mu + iT_7^\mu = \frac{i}{2} s_\theta \left[-D\overline{\hat{K}^0}\left(J_3^{5\mu} + \frac{1}{\sqrt{3}}J_8^{5\mu}\right) + F\hat{K}^0\left(-J_3^{5\mu} + \sqrt{3}J_8^{5\mu}\right)\right];$$

$$T_6^\mu - iT_7^\mu = -\frac{i}{2} s_\theta \left[-D\hat{K}^0\left(J_3^{5\mu} + \frac{1}{\sqrt{3}}J_8^{5\mu}\right) + F\overline{\hat{K}^0}\left(-J_3^{5\mu} + \sqrt{3}J_8^{5\mu}\right)\right];$$

$$T_3^{5\mu} = D\left[J_3^{5\mu} - \frac{1}{2}\left(J_3^{5\mu} + \frac{1}{\sqrt{3}}J_8^{5\mu}\right)s_{\theta/2}^2\right] + F\left[J_3^{5\mu} + \frac{1}{2}\left(-J_3^{5\mu} + \sqrt{3}J_8^{5\mu}\right)s_{\theta/2}^2\right];$$

$$T_8^{5\mu} = \frac{D}{3}\left[-J_8^{5\mu} + \frac{3\sqrt{3}}{2}\left(J_3^{5\mu} + \frac{1}{\sqrt{3}}J_8^{5\mu}\right)s_{\theta/2}^2\right] + F\left[J_8^{5\mu} - \frac{\sqrt{3}}{2}\left(-J_3^{5\mu} + \sqrt{3}J_8^{5\mu}\right)s_{\theta/2}^2\right];$$

$$T_6^{5\mu} + iT_7^{5\mu} = \left[-f_\pi^2 \partial^\mu \theta + \frac{i}{2} s_\theta\left(-J_3^\mu + \sqrt{3}J_8^\mu\right)\right]\overline{\hat{K}^0};$$

$$T_6^{5\mu} - iT_7^{5\mu} = \left[-f_\pi^2 \hat{K}^0 \partial^\mu \theta - \frac{i}{2} s_\theta\left(-J_3^\mu + \sqrt{3}J_8^\mu\right)\right]\hat{K}^0; \tag{A4.7}$$

The additional four vector and four axial vector currents mediate (via $J_\pm^\mu, J_\pm^{5\mu}$) transitions between (proton even, neutron even) states and (proton odd, neutron odd) states: that physics also lies beyond the scope of this paper. Still, for purposes of completeness, we list them:

$$T_1^\mu \pm iT_2^\mu = J_\pm^\mu c_{\theta/2}$$

$$T_4^\mu + iT_5^\mu = \frac{i}{2} s_{\theta/2} \left[D\overline{\hat{K}^0} J_+^{5\mu} + F\hat{K}^0 J_-^{5\mu}\right]$$

$$T_4^\mu - iT_5^\mu = -\frac{i}{2} s_{\theta/2} \left[D\hat{K}^0 J_-^{5\mu} + F\overline{\hat{K}^0} J_+^{5\mu}\right] \tag{A4.8}$$

$$T_1^{5\mu} \pm iT_2^{5\mu} = (F+D)J_\pm^{5\mu} c_{\theta/2}$$

$$T_4^{5\mu} + iT_5^{5\mu} = is_{\theta/2} J_+^\mu \overline{\hat{K}^0}; \quad T_4^{5\mu} - iT_5^{5\mu} = -is_{\theta/2} J_-^\mu \hat{K}^0;$$

## Appendix 5: Further results to order $\Lambda^0_{\chi SB}$ up to linear in $m_S$

Figures 4, 5, A5.1-4 neglect higher $\Lambda^{-1}_{\chi SB}$ order terms (i.e. $C^S_{201}, \overline{C^S_{201}} = 0$). Figure A5.1 and A5.2 show that (to order $\Lambda^0_{\chi SB}$) $\infty S\chi NL$ (typically) has very high baryon number density and chemical potential

$$[\mu^n_B]_{\infty S\chi NL} \geq 0.896 GeV; \quad \frac{[n^\dagger n]_{\infty S\chi NL}}{[\Psi^\dagger \Psi]_{NuclearMater}} \sim 4.84 - 6; \quad (A5.1)$$

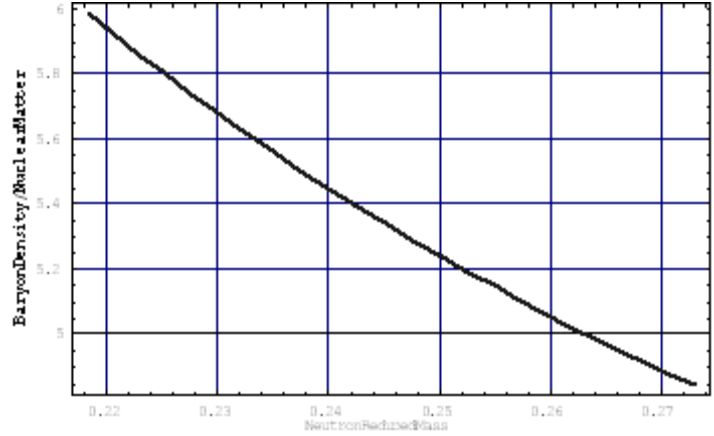

Figure A5.1: $[n^\dagger n]_{\infty S\chi NL}$ vs $[m^n_*]_{\infty S\chi NL}$ to order $\Lambda^0_{\chi SB}$ up to linear in $m_S$ See text for details.

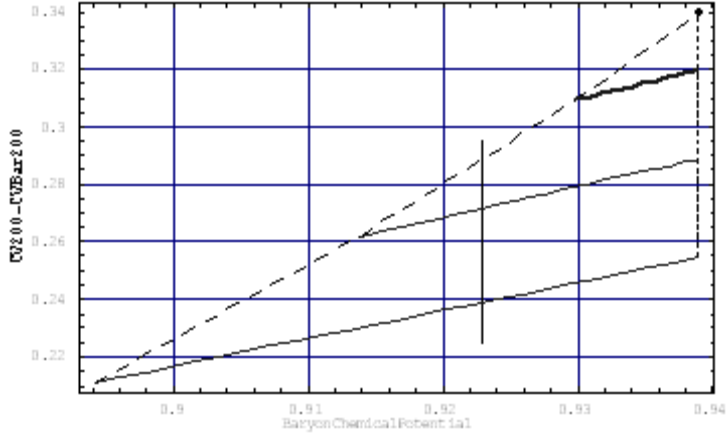

Figure A5.2: $C^V_{200} - \overline{C^V_{200}}$ vs $[\mu^n_B]_{\infty S\chi NL}(GeV)$ to order $\Lambda^0_{\chi SB}$ and linear $m_S$. See text for details. Macroscopic Q-Balls to the left of the vertical solid line have binding energy per nucleon deeper than ordinary nuclear liquids.

The solid lines in Figures 4,5, $A5.2$, $A5.3$ are lines (71) of constant baryon number density $\frac{[n^\dagger n]_{\infty S\chi NL}}{[\Psi^\dagger \Psi]_{NuclearMater}} = 5.990, 5.492, 5.086;$ with increasing thickness respectively. The big dot has $\frac{[n^\dagger n]_{\infty S\chi NL}}{[\Psi^\dagger \Psi]_{NuclearMater}} = 4.840$ and $[\mu^n_*]_{\infty S\chi NL} = m^N$. The dashed and dotted lines are explained in Section 6.3. Physical pairs $(C^S_{200} - \overline{C^S_{200}}, C^V_{200} - \overline{C^V_{200}})$ are defined when self-bound Q-Ball non-topological solitons exist and lie in the (rough) triangle subtended by the dotted, dashed and thinnest solid lines. The dashed lines minimize baryon number density for a given chemical potential. Figure A5.3 demonstrates the mathematical self-consistency condition that $[\theta]_{\infty S\chi NL}$ is indeed an angle.

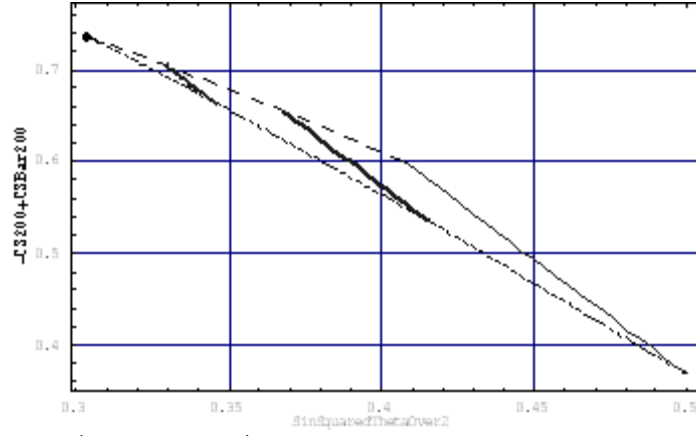

<u>Figure A5.3</u>: $\left(-C_{200}^S + \overline{C_{200}^S}\right)$ vs. $[s_{\theta/2}^2]_{\infty S\chi NL}$ to order $\Lambda_{\chi SB}^0$ and linear $m_S$. See text for details.

## Appendix 6: Algebraic version of previous graphical [14] upper limit on $(-C_{200}^S + \overline{C_{200}^S})$, lower limit on $(C_{200}^V - \overline{C_{200}^V})$ for given $[n^\dagger n]_{\infty S\chi NL}$

For completeness and convenience, we include the higher order $\Lambda_{\chi SB}^{-1}$ terms from Section 6.4 and Appendix 8 explicitly in our expressions here and in the other Appendices (except Appendix 5). Define

$$\hat{\beta}^{nK}\sigma_{\pi N} = \beta^{nK}\sigma_{\pi N} - 2\frac{\left(C_{201}^S - \overline{C_{201}^S}\right)}{f_\pi^2}\frac{m_S}{\Lambda_{\chi SB}}(\bar{n}n); \quad (A6.1)$$

and insist that

$$2\hat{\beta}^{nK}\sigma_{\pi N}\frac{m_*^n}{k_F^n}\frac{ds_{\theta/2}^2}{dk_F^n} \geq 0; \quad 2\hat{\beta}^{nK}\sigma_{\pi N}\frac{m_*^n}{k_F^n}\frac{ds_{\theta/2}^2}{dk_F^n} \xrightarrow[k_F^n \to 0]{} 1; \quad (A6.2)$$

Our algebraic version (applicable to microscopic and macroscopic $S\chi NL$) requires the existence of solutions $(\mu_B^n)_{Minimum}, (k_F^n)_{Minimum}$ such that

$$\left[2\hat{\beta}^{nK}\sigma_{\pi N}\frac{m_*^n}{k_F^n}\frac{ds_{\theta/2}^2}{dk_F^n}\right]_{(\mu_B^n)_{Minimum}, (k_F^n)_{Minimum}} = 0; \quad (A6.3)$$

$$\left[\frac{d}{dk_F^n}\left(2\hat{\beta}^{nK}\sigma_{\pi N}\frac{m_*^n}{k_F^n}\frac{ds_{\theta/2}^2}{dk_F^n}\right)\right]_{(\mu_B^n)_{Minimum}, (k_F^n)_{Minimum}} = 0; \quad (A6.4)$$

and that $\mu_B^n \geq (\mu_B^n)_{Minimum}$. These equations are easily solved. First write $(n^\dagger n, \mu_*^n, m_*^n, \bar{n}n, \theta)$ as functions of $(k_F^n, \mu_B^n)$

$$\mu_*^n = \mu_B^n - \frac{\left(C_{200}^V - \overline{C_{200}^V}\right)}{f_\pi^2} n^\dagger n; \quad m_*^n = \sqrt{\left(\mu_*^n\right)^2 - \left(k_F^n\right)^2}$$

(A6.5)

$$\overline{n}n = \frac{m_*^n}{2\pi^2}\left(k_F^n \mu_*^n - \frac{1}{2}\left(m_*^n\right)^2 \ln\left(\frac{\mu_*^n + k_F^n}{\mu_*^n - k_F^n}\right)\right);$$

$$s_{\theta/2}^2 = \frac{1}{2\hat{\beta}^{nK}\sigma_{\pi N}}\left(m^N - m_*^n + \frac{\left(C_{200}^S - \overline{C_{200}^S}\right)}{f_\pi^2}\overline{n}n\right);$$

(A6.6)

Differentiating this last equation (A6.6) we have

$$2\hat{\beta}^{nK}\sigma_{\pi N}\frac{m_*^n}{k_F^n}\frac{ds_{\theta/2}^2}{dk_F^n} = Z_1 + (Z_2 + Z_1 Z_3)Z_4;$$

(A6.7)

$$Z_1 = \left(1 + \frac{(C_{200}^V - \overline{C_{200}^V})}{\pi^2 f_\pi^2}k_F^n \mu_*^n\right); \quad \frac{dZ_1}{dk_F^n} = \frac{(C_{200}^V - \overline{C_{200}^V})}{\pi^2 f_\pi^2}(4\mu_*^n - 3\mu_B^n);$$

$$Z_2 = \frac{(C_{200}^V - \overline{C_{200}^V})}{\pi^4 f_\pi^2}\left((k_F^n \mu_*^n)^2 - (k_F^n)^4\right); \quad \frac{dZ_2}{dk_F^n} = \frac{(C_{200}^V - \overline{C_{200}^V})}{\pi^4 f_\pi^2}\left(2k_F^n \mu_*^n (4\mu_*^n - 3\mu_B^n) - 4(k_F^n)^3\right);$$

$$Z_3 = 3\frac{\overline{n}n}{m_*^n} - \frac{1}{\pi^2}k_F^n \mu_*^n; \quad \frac{dZ_3}{dk_F^n} = \frac{1}{\pi^2}\left(-4\mu_*^n + 3\mu_B^n + \mu_*^n\left(\frac{k_F^n}{m_*^n}\right)^2 Z_1\right) - \frac{k_F^n}{m_*^n}(3Z_2 + 2Z_1 Z_3);$$

$$Z_4 = -\frac{(C_{200}^S - \overline{C_{200}^S})}{f_\pi^2} - 4\frac{(C_{201}^S - \overline{C_{201}^S})}{f_\pi^2}\frac{m_S}{\Lambda_{\chi SB}}s_{\theta/2}^2;$$

$$\frac{dZ_4}{dk_F^n} = -2\frac{\left(C_{201}^S - \overline{C_{201}^S}\right)}{f_\pi^2}\frac{m_S}{\Lambda_{\chi SB}}\frac{k_F^n}{\hat{\beta}^{nK}\sigma_{\pi N}m_*^n}\left(2\hat{\beta}^{nK}\sigma_{\pi N}\frac{m_*^n}{k_F^n}\frac{ds_{\theta/2}^2}{dk_F^n}\right);$$

It is useful to regard a given pair $(C_{200}^S - \overline{C_{200}^S}, C_{200}^V - \overline{C_{200}^V})$ as known functions of $[\mu_B^n]_{\infty S\chi NL}, [k_F^n]_{\infty S\chi NL}$ (see Appendix 8). Figure A6.1 shows (for the line of constant baryon number density $\frac{[n^\dagger n]_{\infty S\chi NL}}{[\Psi^\dagger \Psi]_{NuclearMater}} = 5.492$ in Figures 4,5, A5.2, A5.3 where $C_{201}^S, \overline{C_{201}^S} = 0$) the surface $2\hat{\beta}^{nK}\sigma_{\pi N}\frac{m_*^n}{k_F^n}\frac{ds_{\theta/2}^2}{dk_F^n}$ as a function of $\mu_B^n, k_F^n$, but evaluated at $\mu_B^n = [\mu_B^n]_{\infty S\chi NL}$.

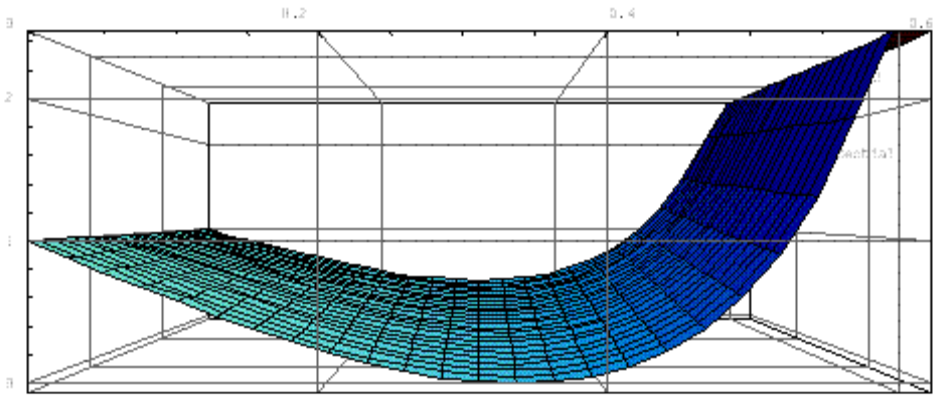

<u>Figure A6.1</u>  $2\hat{\beta}^{nK}\sigma_{\pi N}\frac{m_*^n}{k_F^n}\frac{ds_{\theta/2}^2}{dk_F^n}$ as a function of $k_F^n$ and $\mu_B^n = [\mu_B^n]_{\infty S\chi NL}$ to order $\Lambda_{\chi SB}^0$ and linear $m_S$. $\frac{[n^\dagger n]_{\infty S\chi NL}}{[\Psi^\dagger\Psi]_{NuclearMatter}} = 5.492$. See text for details.

On the surface, Newtonian potential motion is well defined $2\hat{\beta}^{nK}\sigma_{\pi N}\frac{m_*^n}{k_F^n}\frac{ds_{\theta/2}^2}{dk_F^n} \geq 0$, $[k_F^n]_{\infty \chi SNL} \geq ([k_F^n]_{\infty \chi SNL})_{Minimum}$, non-topological solitons exist, and the liquid can't disperse energetically to free neutrons. The solution

$$[\mu_B^n]_{\infty \chi SNL} = ([\mu_B^n]_{\infty \chi SNL})_{Minimum}; \quad k_F^n = (k_F^n)_{Minimum}; \quad 2\hat{\beta}^{nK}\sigma_{\pi N}\frac{m_*^n}{k_F^n}\frac{ds_{\theta/2}^2}{dk_F^n} = 0;$$

$$\frac{d}{dk_F^n}\left(2\hat{\beta}^{nK}\sigma_{\pi N}\frac{m_*^n}{k_F^n}\frac{ds_{\theta/2}^2}{dk_F^n}\right) = 0; \quad (A6.8)$$

is the lowest point on the near edge of the curved surface. The "physical" region
$$[\mu_B^n]_{\infty \chi SNL} \geq ([\mu_B^n]_{\infty \chi SNL})_{Minimum}; \quad (A6.9)$$

defines the dashed lines in Figures 4,5, $A5.2$, $A5.3$ (where $C_{201}^S, \overline{C_{201}^S} = 0$) and Figures 6,7, $A8.2$, $A8.3$ (where $C_{201}^S, \overline{C_{201}^S} \neq 0$).

## **Appendix 7: Higher Order $\theta = 0$ Terms in Skyrme Mean Field Models of Ordinary Heavy Nuclei**

Careful and successful comparison of theory vs. experiment for the ground state of even-even spin-zero spherical closed-shell heavy nuclei is a major triumph for Relativistic Mean Field Point Coupling Hartree-Fock (RMF-PC-HF) "Skyrme" models of nuclear many-body forces [9,12,11]. We showed in Section 5 that Hartree treatment of their exchange terms is equivalent to Hartree-Fock treatment of nuclear states. Their effective Lagrangians are derivable from $SU2 \times U1\chi^{PT}$ chiral liquids [8].

We quote a recent high accuracy fit and its predictions for properties of such heavy nuclei. T.Burvenich, D.G.Madland, J.A.Maruhn & P.-C.Reinhard [9] use 11 coupling constants which, when appropriately rescaled with $\Lambda_{\chi SB}$, (almost) obey naturalness for $SU2 \times U1\chi^{PT}$ chiral nuclear liquids [8]:

1. 2-nucleon forces $\sim \Lambda_{\chi SB}^0$ include exchange terms: coefficients $C_{200}^S, \overline{C_{200}^S}, C_{200}^V, \overline{C_{200}^V}$

$$L^{4f} = -\frac{1}{2}\alpha_S(\overline{\Psi}\Psi)(\overline{\Psi}\Psi) - \frac{1}{2}\alpha_V(\Psi^\dagger\Psi)(\Psi^\dagger\Psi) \\ -\frac{1}{2}\alpha_{TS}(\overline{\Psi}2t_3\Psi)(\overline{\Psi}2t_3\Psi) - \frac{1}{2}\alpha_{TV}(\Psi^\dagger 2t_3\Psi)(\Psi^\dagger 2t_3\Psi); \quad (A7.1)$$

2. Nucleon exchange terms are neglected in their higher order point coupling $L^{hot} = L^{6f} + L^{8f}$ and must be added to preserve quantum loop power counting:

- 3-nucleon forces $\Lambda_{\chi SB}^{-1}$ are smaller than 2-nucleon $\Lambda_{\chi SB}^{0}$ forces [5]. Including both $C_{300}^{S}, C_{300}^{V}$ would over-count independent chiral coefficients. When dropping $\Lambda_{\chi SB}^{-3}$ terms and writing in terms of the baryon number density

$$L^{6f} \to -\frac{1}{3}\beta_S (\Psi^\dagger \Psi)(\Psi^\dagger \Psi)(\Psi^\dagger \Psi); \qquad (A7.2)$$

- 4-nucleon $\Lambda_{\chi SB}^{-2}$ forces are smaller than 3-nucleon $\Lambda_{\chi SB}^{-1}$ forces [5]. Including both $C_{400}^{S}, C_{400}^{V}$ over-counts independent chiral coefficients. When dropping $\Lambda_{\chi SB}^{-4}$ terms and writing in terms of the baryon number density

$$L^{8f} \to -\frac{1}{4}(\gamma_S + \gamma_V)(\Psi^\dagger \Psi)(\Psi^\dagger \Psi)(\Psi^\dagger \Psi)(\Psi^\dagger \Psi); \qquad (A7.3)$$

3. Their nuclear surface $\Lambda_{\chi SB}^{-2}$ terms include nucleon exchange interactions:
   - Including all of $C_{220}^{S}, \overline{C_{220}^{S}}, C_{220}^{V}, \overline{C_{220}^{V}}$ over-counts independent chiral coefficients. When dropping $\Lambda_{\chi SB}^{-4}$ terms and writing in terms of the baryon number density:

$$\begin{aligned}L^{der} \to &-\frac{1}{2}(\delta_S + \delta_V)[\partial_\mu (\Psi^\dagger \Psi)][\partial^\mu (\Psi^\dagger \Psi)] \\ &-\frac{1}{2}(\delta_{TS} + \delta_{TV})[\partial_\mu (\Psi^\dagger 2t_3 \Psi)][\partial^\mu (\Psi^\dagger 2t_3 \Psi)]\end{aligned} \qquad (A7.4)$$

   - Because they only involve differentials of the baryon number density, these surface terms are invariant under local $SU3\chi^{PT}$ transformations, do not contribute to $SU3_{L+R}$ or $SU3_{L+R}$ currents or affect CVC or PCAC. In terms of the baryon and pseudo-Goldstone octets:

$$\begin{aligned}f_\pi^2 \Lambda_{\chi SB}^2 L_{\chi PT}^{4-B;Surface,Symmetric} &\sim Tr([D_\mu(\bar{B}\gamma^A B)][D^\mu(\bar{B}\gamma_A B)]), \; Tr(D_\mu(\bar{B}\gamma^A B))Tr(D^\mu(\bar{B}\gamma_A B)); \\ D_\mu(\bar{B}\gamma^A B) &= \partial_\mu(\bar{B}\gamma^A B) + [V_\mu, \bar{B}\gamma^A B]; \\ D_\mu(\bar{B}\gamma^A B) &\to U(D_\mu(\bar{B}\gamma^A B))U^\dagger;\end{aligned} \qquad (A7.5)$$

$$\begin{aligned}L_{\chi PT;Liquid}^{4-\Psi;Surface,Symmetric} = &-\frac{C_{220}}{f_\pi^2 \Lambda_{\chi SB}^2}[\partial_\mu(\Psi^\dagger \Psi)][\partial^\mu(\Psi^\dagger \Psi)] \\ &+\frac{\overline{C_{220}}}{2f_\pi^2 \Lambda_{\chi SB}^2}([\partial_\mu(\Psi^\dagger \Psi)][\partial^\mu(\Psi^\dagger \Psi)] + [\partial_\mu(\Psi^\dagger 2t_3 \Psi)][\partial^\mu(\Psi^\dagger 2t_3 \Psi)])\end{aligned} \qquad (A7.6)$$

   - These replace the scalar $\sigma$ sigma particle in the nuclear surface [21];

It is beyond the scope of this paper to construct a complete minimal (order $\Lambda_{\chi SB}^{-2}$) set of chiral liquid operators (with $\theta = 0$) for nuclear Skyrme models, but a systematic program of calculation of detailed properties of the ground state of even-even spin-zero spherical closed-shell heavy nuclei in RMF-PC-HF (and ordinary nuclear liquid drops) with that set is necessary in order to extract predictions for nuclear structure from $SU2 \times U1\chi^{PT}$. According to strict order $\Lambda_{\chi SB}^{-2}$ power counting [5], current nuclear Skyrme models:

- Sometimes double-count chiral liquid operators: e.g. to order $\Lambda_{\chi SB}^{-2}$, only eight of 11 coupling constants above are independent;
- Are missing many chiral liquid operators: e.g. although, the effect of operator

$$\frac{1}{f_\pi^2 \Lambda_{\chi SB}} Tr\big([\bar{B}\gamma^A D_\mu \gamma^\mu B][\bar{B}\gamma_A B]\big) \to \frac{1}{f_\pi^2 \Lambda_{\chi SB}} [\bar{\Psi}\partial_\mu \gamma^\mu \Psi][\bar{\Psi}\Psi]; \quad (A7.7)$$

on $S\chi NL$ baryon density and chemical potential is small [14], it may be important for high accuracy nuclear structure. But this operator will also affect ordinary heavy nuclei, $SU3_{L+R}$ and $SU3_{L+R}$ currents, CVC and PCAC.

Going forward, a new round of high accuracy nuclear experiments can place better constraints on chiral coefficients. Crucial among these are measurements of the neutron density in $^{82}_{126}Pb_{208}$ [40], which can better fix $\left(C^S_{200} - \overline{C^S_{200}}, C^V_{200} - \overline{C^V_{200}}\right)$. The results will also place strong constraints on $S\chi NL$.

**Appendix 8: Higher Order $\theta \neq 0$ Symmetry Breaking Terms $\sim m_S$ in $\infty S\chi NL$**

In Section 6.4, the Lagrangian of Appendix 1 is supplemented by (higher order $\Lambda_{\chi SB}^{-1}$ linear in $m_S$) explicit chiral symmetry breaking 4-fermion point-coupling terms which do not contribute to vector or axial vector currents:

$$L_{\chi PT}^{\pi,4-B;SymmetryBreaking} \to L_{\chi PT}^{K,4-\Psi;SymmetryBreaking;Exchange}$$

$$\xrightarrow[NeutralS\chi NL]{} -2\frac{\left(C^S_{201} - \overline{C^S_{201}}\right)}{f_\pi^2}\frac{m_S}{\Lambda_{\chi SB}} s^2_{\theta/2}(\bar{n}n)(\bar{n}n); \quad (A8.1)$$

All formulae in the Appendices (except Appendix 5) have been modified to include the effect of this term. We summarize various results here and in Section 6.4. Define

$$\left[\hat{\beta}^{nK}\sigma_{\pi N} - \beta^{nK}\sigma_{\pi N} + 2\frac{\left(C^S_{201} - \overline{C^S_{201}}\right)}{f_\pi^2}\frac{m_S}{\Lambda_{\chi SB}}(\bar{n}n)\right]_{\infty S\chi NL} = 0; \quad (A8.2)$$

The Conservation of Chiral Axial-vector Currents in $\infty S\chi NL$ equation (60) shows

1. $$\left[\beta^{nK}\sigma_{\pi N}\bar{n}n - f_\pi^2 m_K^2 - \frac{\left(C^S_{201} - \overline{C^S_{201}}\right)}{f_\pi^2}\frac{m_S}{\Lambda_{\chi SB}}(\bar{n}n)(\bar{n}n)\right]_{\infty S\chi NL} = 0; \quad (A8.3)$$

    which can give (for $C^S_{201} - \overline{C^S_{201}} < 0$) a significant reduction in scalar density;

2. The 1to1 relation between $[m_*^n]_{\infty S\chi NL}$ and $[n^\dagger n]_{\infty S\chi NL}$ remains independent of $\left(C^S_{200}, C^V_{200}, \overline{C^S_{200}}, \overline{C^V_{200}}\right)$:

$$[n^\dagger n]_{\infty S\chi NL} = \frac{1}{3\pi^2}\left([k_F^n]_{\infty S\chi NL}\right)^3; \quad [\mu_*^n]_{\infty S\chi NL} = \left[\sqrt{(k_F^n)^2 + (m_*^n)^2}\right]_{\infty S\chi NL};$$

$$[\bar{n}n]_{\infty S\chi NL} = \left[\frac{m_*^n}{2\pi^2}\left(k_F^n \mu_*^n - \frac{1}{2}(m_*^n)^2 \ln\left(\frac{\mu_*^n + k_F^n}{\mu_*^n - k_F^n}\right)\right)\right]_{\infty S\chi NL}; \quad (A8.4)$$

3. $C^V_{200} - \overline{C^V_{200}}$ determines a 1to1 relationship between $[\mu_B^n]_{\infty S\chi NL}$ and $[n^\dagger n]_{\infty S\chi NL}$:

$$[\mu_B^n]_{\infty S\chi NL} = \left[\mu_* + \frac{C^V_{200} - \overline{C^V_{200}}}{f_\pi^2}n^\dagger n\right]_{\infty S\chi NL}; \quad (A8.5)$$

4. A 1to1 relationship between $[s^2_{\theta/2}]_{\infty S\chi NL}$ and $[n^\dagger n]_{\infty S\chi NL}$ is determined by $(C^S_{200} - \overline{C^S_{200}}, C^S_{201} - \overline{C^S_{201}})$:

$$[m^n_*]_{\infty S\chi NL} = m^N + \left[\frac{(C^S_{200} - \overline{C^S_{200}})}{f^2_\pi}\bar{n}n - 2\hat{\beta}^{nK}\sigma_{\pi N}s^2_{\theta/2}\right]_{\infty S\chi NL} ; \qquad (A8.6)$$

The Zero Total $\infty S\chi NL$ Pressure equation (59)

5. Gives another 1to1 relationship between $[\mu^n_B]_{\infty S\chi NL}$ and $[n^\dagger n]_{\infty S\chi NL}$, this time determined by $(C^S_{200} - \overline{C^S_{200}}, C^S_{201} - \overline{C^S_{201}})$:

$$\left[\left(\frac{1}{2}\mu^n_B - \frac{1}{4}\mu^n_*\right)n^\dagger n - \frac{1}{4}m^n_*\bar{n}n + \frac{(C^S_{200} - \overline{C^S_{200}})}{2f^2_\pi}(\bar{n}n)^2\right]_{\infty S\chi NL}$$
$$- 2\left[\left(m^2_K f^2_\pi - \frac{(C^S_{201} - \overline{C^S_{201}})}{f^2_\pi}\frac{m_S}{\Lambda_{\chi SB}}(\bar{n}n)^2\right)s^2_{\theta/2}\right]_{\infty S\chi NL} = 0; \qquad (A8.7)$$

An extreme case $(a_3, C^S_{201} - \overline{C^S_{201}}) = (1.5, -1.5)$, corresponding to Reference [14]'s $(I_S, I_V, \overline{C^S_{201}}) = (3,0,0)$, is instructive. Figure A8.1 and A8.1 show significant reduction in baryon number density and chemical potential (compare Equation 68):

$$[\mu^n_B]_{\infty S\chi NL} \geq 0.807 GeV; \quad \frac{[n^\dagger n]_{\infty S\chi NL}}{[\Psi^\dagger \Psi]_{NuclearMater}} \sim 3.36 - 6; \qquad (A8.8)$$

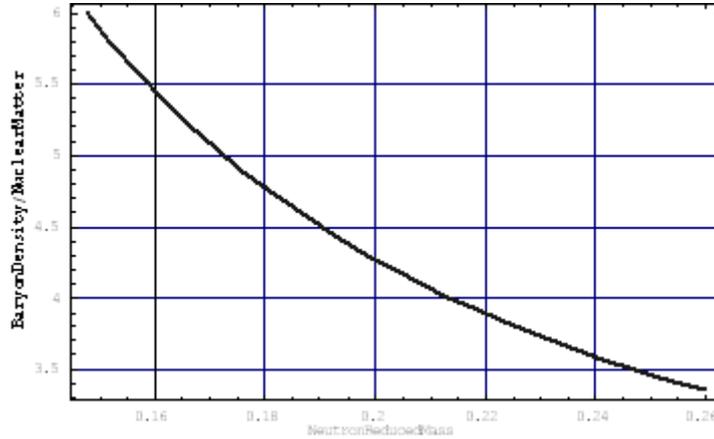

Figure A8.1: $[n^\dagger n]_{\infty S\chi NL}$ vs $[m^n_*]_{\infty S\chi NL}$ to order $\Lambda^{-1}_{\chi SB}$ and linear in $m_S$ with $C^S_{201} - \overline{C^S_{201}} = -1.5$ and $a_3 = 1.5$. See text for details.

The slanted solid lines in Figures $A8.2, A8.3$ are of constant baryon number density $\frac{[n^\dagger n]_{\infty S\chi NL}}{[\Psi^\dagger \Psi]_{NuclearMater}} = 6.00, 4.00, 3.66;$ with increasing thickness (compare Equation 71). The big dot has $\frac{[n^\dagger n]_{\infty S\chi NL}}{[\Psi^\dagger \Psi]_{NuclearMater}} = 3.36$ and $[\mu^n_*]_{\infty S\chi NL} = m^N$ The dashed and dotted lines are explained in Section 6.3. Physical pairs $(C^S_{200} - \overline{C^S_{200}}, C^V_{200} - \overline{C^V_{200}})$ are defined when self-bound Q-Ball non-topological solitons exist and lie in the area subtended by the dotted, dashed and thinnest solid lines. The dashed lines minimize baryon number density for a given chemical potential. Figure A8.2 shows that baryon number densities and chemical potentials can also be much lower (compare Figure A5.2).

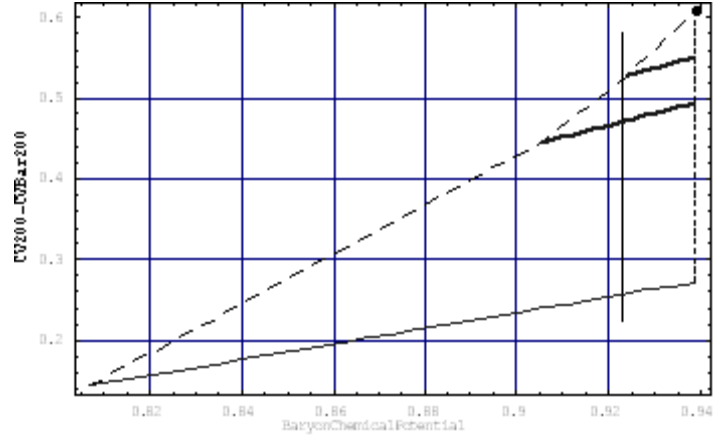

Figure A8.2: $C_{200}^V - \overline{C_{200}^V}$ vs $[\mu_B^n]_{\infty S\chi NL}(GeV)$ to order $\Lambda_{\chi SB}^{-1}$ and linear $m_S$ with $C_{201}^S - \overline{C_{201}^S} = -1.5$ and $a_3 = 1.5$. See text for details. Macroscopic Q-Balls to the left of the vertical solid line have binding energy per nucleon deeper than ordinary nuclear liquids.

Figure A8.3 verifies that $[\theta]_{\infty S\chi NL}$ is an angle.

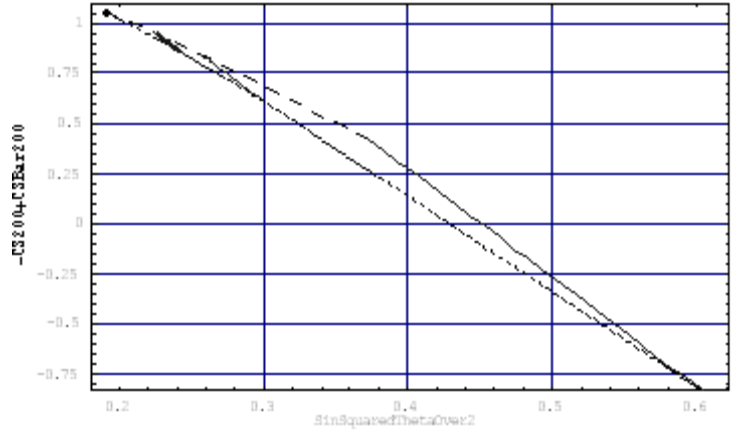

Figure A8.3: $\left(-C_{200}^S + \overline{C_{200}^S}\right)$ vs $[s_{\theta/2}^2]_{\infty S\chi NL}$ to order $\Lambda_{\chi SB}^{-1}$ and linear $m_S$ with $C_{201}^S - \overline{C_{201}^S} = -1.5$ and $a_3 = 1.5$. See text for details.

**A8.2 New lower $(-C_{200}^S + \overline{C_{200}^S})$ limit, upper $(C_{200}^V - \overline{C_{200}^V})$ limit for given $[n^\dagger n]_{\infty S\chi NL}$**

Figure A8.4 shows that $2\hat{\beta}^{nK}\sigma_{\pi N}\dfrac{m_*^n}{k_F^n}\dfrac{ds_{\theta/2}^2}{dk_F^n}$ vs. $k_F^n, [\mu_B^n]_{\infty S\chi NL}$ turns negative toward the right hand side, giving a new upper bound $\left([\mu_B^n]_{\infty \chi SNL}\right)_{Maximum}$ from the requirement that (everywhere along the path of Newtonian potential motion) $2\hat{\beta}^{nK}\sigma_{\pi N}\dfrac{m_*^n}{k_F^n}\dfrac{ds_{\theta/2}^2}{dk_F^n} \geq 0$

$$[\mu_B^n]_{\infty \chi SNL} \leq \left([\mu_B^n]_{\infty \chi SNL}\right)_{Maximum} \approx 0.92 GeV; \qquad (A8.9)$$

This further limits the "physical" region $(C_{200}^S - \overline{C_{200}^S}, C_{200}^V - \overline{C_{200}^V}, C_{201}^S - \overline{C_{201}^S})$ of chiral coefficients where Q-Balls exist.

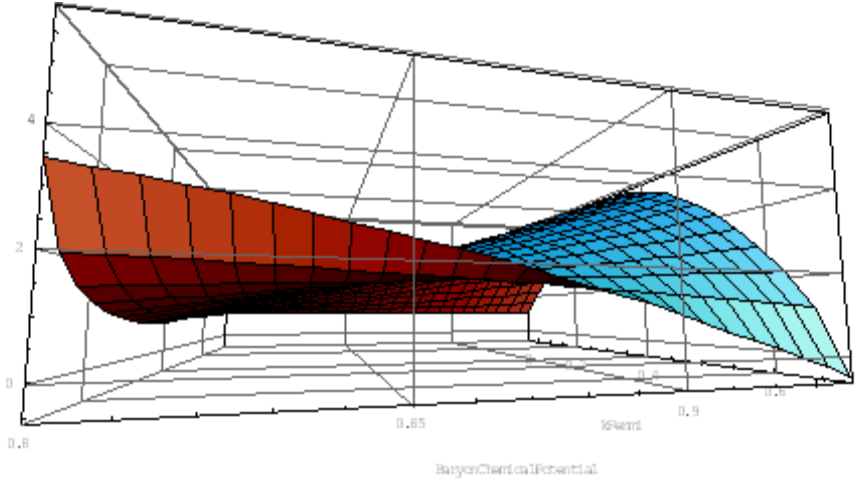

<u>Figure A8.4</u>: $2\hat{\beta}^{nK}\sigma_{\pi N}\frac{m_*^n}{k_F^n}\frac{ds_{\theta/2}^2}{dk_F^n}$ as a function of $k_F^n$ and $\mu_B^n = [\mu_B^n]_{\infty S\chi NL}$ to order $\Lambda_{\chi SB}^{-1}$ and linear $m_S$ with $C_{201}^S - \overline{C_{201}^S} = -1.5$ and $a_3 = 1.5$. Baryon number density $\frac{[n^\dagger n]_{\infty S\chi NL}}{[\Psi^\dagger \Psi]_{NuclearMatter}} = 7$. See text for details.